\documentclass[journal,twocolumn]{IEEEtran}
\usepackage{mathrsfs}
\usepackage{latexsym}
\usepackage{epsfig}
\usepackage{epstopdf}
\usepackage{subfigure} %子图横向排列
\usepackage{array}
\usepackage{amsmath}
\usepackage{amssymb}
\usepackage{color, soul}
\usepackage{amsthm}
\usepackage{enumerate}
\usepackage{algpseudocode}
\usepackage{algorithm}
\usepackage{bm}
\usepackage{balance}
\usepackage[noadjust]{cite}
\usepackage{float}
\usepackage{graphicx}
\usepackage{colortbl}
\usepackage{flushend}
\usepackage{graphics}
\usepackage{cite}
\usepackage{url}

\theoremstyle{plain} %default

\newtheorem{lemma}{Lemma}

\theoremstyle{definition}

\def\bal#1\eal{\begin{align}#1\end{align}}

\newcommand{\bp} {\begin{proof}}
\newcommand{\ep} {\end{proof}}

\newcommand{{\Rb}} {\right)}

\newcommand{{\Rf}} {\right\}}
\begin{document}

%\title{Joint Secure Beamforming and Trajectory Optimization for Active Intelligent Reflecting Surface Assisted Jittering UAV}
\title{Active Intelligent Reflecting Surface Assisted Secure Air-to-Ground Communication with UAV Jittering}
\author{Yimeng Ge and Jiancun Fan, \emph{Senior Member, IEEE}
 \vspace*{-0.2in}

\thanks{This work is partially supported by the National Natural Science Foundation of China under Grants No. 61671367, the Key Research and Development Plan of Shaanxi Province under Grant No. 2018GY-003, the Research Foundation of Science and Technology on Communication Networks Laboratory, and the Fundamental Research Funds for the Central Universities. (\textit{Corresponding author: Jiancun Fan}.)}
\thanks{Y. Ge and J. Fan are with the School of Information and Communication Engineering, Xi'an Jiaotong University, Xi'an, Shaanxi 710049, P. R. China (e-mail: gym0415@stu.xjtu.edu.cn;fanjc0114@gmail.com)}
\thanks{Copyright (c) 2015 IEEE. Personal use of this material is permitted. However, permission to use this material for any other purposes must be obtained from the IEEE by sending a request to pubs-permissions@ieee.org.}
}

\maketitle
 \pagenumbering{gobble}
%======================================================================
\begin{abstract}
Unmanned Aerial Vehicles (UAV)-enabled communication is a promising solution for secure air-to-ground (A2G) networks due to the additional secure degrees of freedom afforded by mobility. However, the jittering characteristics caused by the random airflow and the body vibration of the UAV itself have a non-negligible impact on the performance of UAV communication. Considering the impact of UAV jittering, this paper propose a robust and secure transmission design assisted by an novel active intelligent reflecting surface (IRS), where the reflecting elements in IRS not only adjust the phase shift but also amplify the amplitude of signals. Specifically, under the worst-case secrecy rate constraints, we aim to minimize the transmission power by the robust joint design of active IRS's reflecting coefficient and beamforming at the UAV-borne base station (UBS). However, it is challenging to solve the joint optimization problem due to its non-convexity. To tackle this problem, the non-convex problem is reformulated with linear approximation for the channel variations and linear matrix inequality transformed by S-procedure and Schur's complement. Then, we decouple this problem into two sub-problems, namely, passive beamforming and active IRS's reflecting coefficient optimization, and solve them through alternate optimization (AO). Finally, the numerical results demonstrate the potential of active IRS on power saving under secure transmission constraints and the impact of UAV jittering.

\end{abstract}

\begin{IEEEkeywords}
active intelligent reflecting surface, secure UAV communications, robust beamforming, worst-case model.
\end{IEEEkeywords}

%======================================================================
\section{Introduction}
Unmanned aerial vehicle (UAV) communication has received widespread attention due to its low cost, high flexibility and mobility, which will play an important role in the fifth-generation (5G) networks \cite{UAV_5G}. Compared with traditional terrestrial communications, air-to-ground (A2G) channels are usually dominated by line of sight (LOS) links, which helps to establish long-range coverage and reliable transmission \cite{UAV_LOS}. In addition, the UAV can fly close to its intended ground users to obtain better channel conditions. Despite the above benefits, A2G channels are vulnerable to various attacks, especially eavesdropping attacks from illegal eavesdroppers \cite{UAV_secure}. In this regard, the physical layer security (PLS) technology is considered as a promising method to ensure secure UAV communication. Most of the related researches jointly optimize the UAV trajectory and resource allocation to improve the secure communication quality. Among them, one type of effective cooperative jamming method is to prevent eavesdropping by using the interference of other nearby UAV to ensure the safety of unmanned aerial vehicle (UAV) communications \cite{UAV_PLS_1,UAV_PLS_2,UAV_PLS_3,UAV_PLS_4,Deep_UAV_noIRS}, where one UAV transmitter delivers the confidential information to a ground node, and the other UAV jammer cooperatively sends the artificial noise to confuse the ground eavesdropper. In particular, the secure UAV communication in different scenarios was studied in \cite{UAV_PLS_MEC,UAV_PLS_Vehicle,UAV_PLS_NOMA,UAV_PLS_CR}. A secure UAV-enabled mobile edge computing (MEC) system has been proposed in the presence of multiple eavesdropping UAVs with imperfect locations \cite{UAV_PLS_MEC}. Then, in \cite{UAV_PLS_Vehicle} the authors investigates the secrecy performance of a UAV-to-vehicle communication system. Furthermore, a secure downlink multi-user transmission scheme enabled by a flexible UAV borne base station (BS) and non-orthogonal multiple access (NOMA) was rigorously studied in \cite{UAV_PLS_NOMA}. In \cite{UAV_PLS_CR}, a secure UAV-enabled cognitive radio network is studied by robustly optimizing the UAV's trajectory and transmit power. However, the secrecy performance in these UAV communications is still unsatisfactory because all the above methods are essentially unable to improve the secure degrees of freedom \cite{SDOF}.

In this situation, intelligent reflecting surface (IRS) as a promising technology has attracted great attention. By independently controlling passive reflective elements on the plane, it can produce a desired phase shift on the impinging electromagnetic waves \cite{DEEP_IRS}, thereby providing more secure degrees of freedom for the wireless link. Furthermore, each reflected signals can be added constructively at the legitimate receiver to enhance its reception quality, while being received destructively at eavesdropper to degrade its signal reception \cite{FullDuplex_IRS}. This indicates IRS is an excellent solution to significantly improve the PLS performance \cite{IRS_UAV}. Thus, the design of IRS-assisted secure UAV communication systems has attracted increasing attention. In \cite{UAV_IRS_1}, a theoretical framework was presented to analyze the performance of an integrated UAV-IRS relaying system. Furthermore, in \cite{UAV_IRS_2}, the secure transmission problem was considered in an UAV and IRS assisted millimeter wave (mmWave) networks in the presence of an eavesdropper. In addition, \cite{IRS_UAV_mmWave} showed that the use of IRSs can effectively improve the coverage and reliability of mmWave UAV communication systems. In fact, all these results indicate that IRS greatly boosts the UAV transmission performance compared with no IRS case.

\subsection{Backround and Motivation}
Although the integration of IRS technology with UAV secure communication is a great step toward the realization of efficient and flexible aerial networks. It is observed that among the current related works, there exists many bottlenecks to be overcome on the design of secure UAV communication. How to fully exploit the potential of the IRS and make it applicable to practical scenarios to further improve the secrecy performance of UAVs is an important issue. Especially in multi-UAV networks, the communication nodes are usually scattered in a relatively wide geographic range and long-distance transmission will bring more serious challenges.

In fact, all these aforementioned existing works have ignored an unavoidable problem in IRS-assisted UAV communication: Although IRS brings new reliable reflection link for signal transmission in addition to the direct link, a ``double fading" effect always exists in this reflection link, i.e., the signals received via this link suffer from large-scale fading twice \cite{Active_Passive}. And if the fading coefficient is large, the signals from this longer reflection link lose more power than that from the short direct link, resulting in a limited secrecy performance gain compared with the one without IRS. Recently, a new concept of active IRS has been proposed \cite{Active_Passive,Active_IRS,Active_IRS_5,Active_IRS_2,Active_IRS_3} which can combat this ``double fading" effect. Compared with passive IRS, the active IRS directly reflects the incident signal with the power amplification in the electromagnetic (EM) level, and it still takes the advantage of IRS that no complex and power-hungry radio frequency (RF) chain components are needed. Specifically, in \cite{Active_Passive,Active_IRS,Active_IRS_5}, an active IRS assisted single-input multiple-output uplink system and downlink multiple-input single-output multi-user system were studied respectively, and numerical algorithms were developed to optimize the phase shift and amplification factor of active IRS. Based on the numerical examples, this active IRS design can realize a significantly higher transmission rate compared with those via passive IRS, thus overcoming the fundamental limit of ``double fading" effect.

Essentially, the key of active IRS is that after the incident signal that pass through the BS-IRS subchannel, it typically has a weaker signal power and can be easily amplified with a considerable gain at the expense of low power consumption. Therefore, the physical size of the active IRS can be reduced, making it more suitable for UAV communication limited by the payload. Furthermore, it offers more freedom to design secure beamforming by optimizing the amplitudes of the reflection coefficients instead of just phases, resulting in more flexibility to reconfigure the wireless propagation environment \cite{Active_IRS_3}, \cite{Active_IRS_4}. As a result, the secure UAV communication enhanced by active IRSs is a promising solution for providing high-efficient communication services. However, all of active IRS study \cite{Active_Passive,Active_IRS,Active_IRS_5,Active_IRS_2,Active_IRS_3} did not consider secure UAV communications. Although active IRS greatly helps enhancing the quality of communications of legitimate user, it also reduces ``double fading" effect in the reflection link of UAV-IRS-eavesdropper due to the broadcast nature of wireless channels, resulting larger information leakage to eavesdropper. Hence, the current proposed algorithms is not applicable in secure UAV communication case, and the transmit beamformer at UAV-borne base station (UBS) as well as active IRS's reflecting coefficient should be carefully re-designed via new efficient algorithms.

On the other hand, the jittering characteristics caused by the random airflow and the body vibration of the UAV itself have a non-negligible impact on the establishment of a robust and secure communication link. Unlike terrestrial cellular networks with fixed and stable infrastructure, UAVs are susceptible to airflow and body vibration, resulting in random vibrations such as yaw jittering in the horizontal direction or pitch jittering in the vertical direction \cite{EE_UAV}. It should also be pointed out that the structure of most BS is not specifically designed and optimized for UAV. The weight of the UBS is large and its center of gravity is often not in the middle of the equipment, which makes it very easy to shake, especially when the UAV is vibrating. As a result, the vibration introduces a non-negligible channel estimation error, which results in the deviation of the directional signal beam, and further causes severe performance loss \cite{Vib_UAV}. It must be pointed out that the equivalent noise due to the channel estimation error will also be amplified by active IRS, which means the robust design considering the impact of UAV jittering is essential in active IRS-assisted UAV secure communication.

However, most existing UAV communication research assumes the perfect channel state information (CSI) for A2G links. Only a few papers have studied UAV communication under imperfect CSI caused by channel estimation error. For example, \cite{UAV_IRS_3} maximized the average worst-case secrecy rate by the robust joint design of the UAV's trajectory, IRS's passive beamforming, and transmission power of the legitimate transmitters under the premise of imperfect CSI. In \cite{Deep_UAV}, a precisely designed neural network is adopted to optimize the beamformer for confidential signal and artificial noise, with partial CSI of legitimate UAV and eavesdropping UAV. In adittion, the authors of \cite{Deep_UAV_2} studied the robust and secure transmission in the mmWave UAV communication assisted by IRS under imperfect CSI. In fact, it can be found from the above research that the secrecy performance of IRS-assisted communication is largely limited by the CSI error, and robust design needs to be considered in all scenarios. Therefore, we also need to consider the impact of CSI error caused by UAV jittering in follow-up research.

\subsection{Contribution}
Motivated by the aforementioned facts, this paper develops a comprehensive mathematical framework to characterize the robust secrecy performance of an active IRS-assisted UAV communication, where the impact of UAV jittering on secure beamforming is also considered. To the best of our knowledge, in most current research on secure UAV communication, maximizing achievable secure rate (ASR) is usually the optimization objective. However, maximizing ASR will only guarantee the legitimate user's information security while ignoring their data rates restriction and the UAV's energy efficiency. In fact, an UAV-enabled system should be designed to satisfy the data-link security requirements of its legitimate users as well as to improve the energy efficiency or reduce transmit power, which drives us to investigate a more general scenario and adapt to some complex scenarios, such as power-constrained UAV and distributed UAV system where the communication nodes are usually scattered in a relatively wide geographic range so that long-distance transmission will bring high power consumption. Therefore, considering the legitimate user data rate restriction and the total power constraints of UBS, we study the transmission power minimization problem for active IRS-assisted UAV communication while ensuring ASR. In the proposed solution, power-constrained UAV can use less transmission power to achieve the same ASR, which greatly improves the power-efficiency. Specifically, we formulated the problem of minimizing UBS transmission power by jointly designing the active IRS's reflecting coefficient and beamforming at the UBS, revealing the positive effect of active IRS on the secure UAV communications and the impact of UAV jittering on beamforming. The main contributions of this paper are summarized as follows:

(1) To the best of our knowledge, this is the first work to consider active IRS-assisted UAV secure communication with consideration of UAV jittering, which is more practical and power-efficient than the previous related works. Note that all the existing numerical solutions cannot be directly applied to this setting. Due to the uncertainty factors of UAV jittering including yaw vibration and pitch vibration, accurate estimates of the CSI are usually not available. Specifically, imperfect estimation of antenna array response is characterized with variations on the elevation angle-of-departure (AOD) and azimuth AOD at UAV. Hence, we assume imperfect CSI acquisition of the channels and use a bounded model \cite{imCSI1} to describe the CSI uncertainty. To make the subsequent optimization problems easier to solve, linear approximation applying by Taylor expansion is used to characterize the impact of the variations of elevation AOD and azimuth AOD on the channel response. On this basis, we formulate a non-convex robust secure beamforming optimization problem to jointly design the transmit precoding and active IRS's reflecting coefficient matrix.

(2) Because the problem involves optimization of non-concave objective functions subject to non-convex constraints, for which the optimal solution is computationally difficult to find. Thus, S-procedure, Schur's complement and slack variables are firstly applied to transform the restrictions into linear matrix inequalities (LMI). Then, the optimization problem is decoupled into two sub-problems: 1) Optimize the active IRS's reflecting coefficient matrix for given beamforming matrix; 2) Optimize the beamforming matrix for given IRS's reflecting coefficient matrix. After that, we solved them effectively through the alternating optimization (AO) techniques, in which each set of variables is alternately optimized. Specifically, S-procedure is adopted to approximate the semi-infinite inequality constraints for the first sub-problem. Subsequently, the second sub-problem of transmit precoding is solved by Schur's complement. Thus, it is a convex optimization problem with a linear objective function and LMI constraints, which can be easily solved by CVX.\\
\indent (3) Finally, numerical results are provided to show the efficiency of the proposed optimization scheme and to demonstrate the impact of active IRS on the power saving. Specifically, compared with passive IRS, the proposed active IRS-assisted scheme can significantly reduce transmitting power at UBS while ensuring ASR, even in the complex situation of UAV's power limitation, severe jittering, and high flight altitude. Besides, the secure UAV communication can get a significant performance improvement by the active IRS with a small number of elements, which is more suitable for drones. It is worth pointing out that unlike passive IRS, the performance gain from active IRS will not continue to improve with the increase of the number of elements and the maximum amplification gain, that is, the best performance requires a reasonable design. Our solution also provides a design basis for the active IRS-assisted UAV communication.\\
\indent The rest part of our paper is organized as follows. Section II introduces the system model including A2G wireless channel Model, jittering UAV model, and signal model. Sections III formulates the worst-case robust beamforming problem to minimize the transmit power, and presents the AO algorithm to solve the problems in the cases of active IRS and passive IRS, respectively. Section VI presents numerical results to evaluate the secrecy performance of the proposed algorithms and finally Section V concludes the paper.\\
\indent \emph{Notations}:
Boldface lowercase and uppercase letters denote vectors and matrices, respectively. For a vector $\mathbf{x}=[x_1,\cdots,x_n]$, diag$(\mathbf{x})$ denotes a diagonal matrix whose entries are $x_1,\cdot\cdot\cdot,x_n$. The symbols $\|\mathbf{X}\|_{F}$ denote Frobenius norm of matrix $\mathbf{X}$ and the symbol $\|\mathbf{x}\|_{2}$ denotes 2-norm of vector $\mathbf{x}$. Moreover, $(\cdot)^{H}$ is the conjugate transpose, $(\cdot)^{T}$ is the transpose, $(\cdot)^{*}$ is the conjugate matrix, and $|\cdot|$ is the absolute value.  $\mathbf{I}$ and $(\cdot)^{-1}$ denote the identity matrix and the inverse of a matrix, respectively. The Kronecker product and Hadamard product are denoted by $\otimes $ and $\odot$, respectively. By $\mathbf{X} \succeq 0$ or $\mathbf{X} \succ \mathbf{0}$, we mean that $\mathbf{X}$ is positive semi-definite or positive definite, respectively. $\mathbb{C}^{x \times y}$ denotes the $x \times y$ domain of complex matrices, $\operatorname{Re}\{\cdot\}$ denotes the real part. $\mathcal{C}\mathcal{N}\left(0, \delta^{2}\right)$ represent the real-valued and complex-valued Gaussian distribution with mean 0 and covariance $\delta^{2}$.
%======================================================================
\section{System Model}
In this section, we establish the A2G wireless channel model, CSI error models considering UAV jittering, and signal model in an active IRS-assisted UAV secure communication system as follows.
\subsection{A2G Wireless Channel Model}
\begin{figure}[htbp]
\centerline{\includegraphics[width=2.2in]{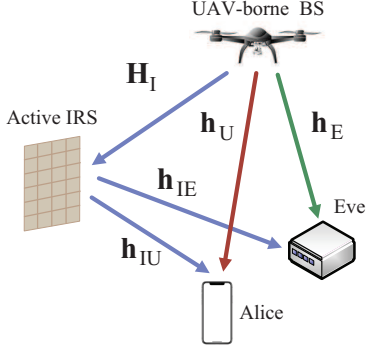}}
	\caption{A2G wiretap model.} \label{scene_dis}
\label{system}
\end{figure}

Consider an active IRS-assisted UAV communication system, as in Fig. \ref{system}, including a single-antenna legitimate users Alice and a single-antenna eavesdropper Eve.
Both the UBS and the active IRS are equipped with uniform rectangular arrays (URAs) with the size of $N=N_{x} \times N_{y} $ and $M=M_{x} \times M_{y}$ respectively, where $N_{x}\left(N_{y}\right)$ and $M_{x}\left(M_{y}\right)$ denote the numbers of UBS antennas and REs along the $x(y)$ axis, respectively. All communication nodes are placed in the three-dimensional (3D) Cartesian coordinate system. The coordinations of UAV, Alice and Eve are $\left(x_{\mathrm{U}}, y_{\mathrm{U}}, z_{\mathrm{U}}\right)$, $\left(x_{\mathrm{A}}, y_{\mathrm{A}}, 0\right)$ and $\left(x_{\mathrm{E}}, y_{\mathrm{E}}, 0\right)$, respectively. The reflecting coefficient matrix for the active IRS is denoted by $\mathbf{\Theta}=\operatorname{diag}\left\{\tau_{1} e^{j \theta_{1}}, \tau_{2} e^{j \theta_{2}}, \ldots, \tau_{M}e^{j \theta_{M}}\right\} \in \mathbb{C}^{M \times M}$, where $\tau_{m}$ and $\theta_{m}$ represents the amplitude and the phase, respectively, and $\tau_{m}$ can be greater than 1 with active load. By contrast, the passive IRS cannot amplify the incident signal, so the amplitude of each RE is limited with $\tau_{m} \leqslant 1$. Furthermore, the first element of the IRS is regarded as the reference point whose coordinates are denoted by $\left(x_{\mathrm{I}}, y_{\mathrm{I}}, z_{\mathrm{I}}\right)$. Therefore, the distance between the IRS and a certain communication node can be approximated by that between the reference point and the corresponding node. We assume that both UBS-Alice/Eve links and UBS-IRS link contain LOS and non-LOS (NLOS) components, which follow the free-space path loss model [19].  Following \cite{UAV_IRS_4}, we consider that all the channels are formulated as the product of large scale fading and small scale fading, and the small scale fading in the A2G links are assumed to be Rician fading. In particular, $\mathbf{h}_{\mathrm{U}}$, $\mathbf{h}_{\mathrm{E}}$ and $\mathbf{H}_{\mathrm{I}}$ can be modeled as
\begin{equation}
\label{channel}
\begin{aligned}
&\mathbf{h}_{\mathrm{U}}=\sqrt{\frac{A_{\mathrm{L}} d_{\mathrm{U}}^{-\alpha_{\mathrm{L}}} K_{\mathrm{U}}}{1+K_{\mathrm{U}}}} \mathbf{h}_{\mathrm{U}, \mathrm{L}}+\sqrt{\frac{A_{\mathrm{N}} d_{\mathrm{U}}^{-\alpha_{\mathrm{N}}}}{1+K_{\mathrm{U}}}} \mathbf{h}_{\mathrm{U}, \mathrm{N}}, \\
&\mathbf{h}_{\mathrm{E}}=\sqrt{\frac{A_{\mathrm{L}} d_{\mathrm{E}}^{-\alpha_{\mathrm{L}}} K_{\mathrm{E}}}{1+K_{\mathrm{E}}}} \mathbf{h}_{\mathrm{E}, \mathrm{L}}+\sqrt{\frac{A_{\mathrm{N}} d_{\mathrm{E}}^{-\alpha_{\mathrm{N}}}}{1+K_{\mathrm{E}}}} \mathbf{h}_{\mathrm{E}, \mathrm{N}}, \\
&\mathbf{H}_{\mathrm{I}}=\sqrt{\frac{A_{\mathrm{L}} d_{\mathrm{I}}^{-\alpha_{\mathrm{L}}} K_{\mathrm{I}}}{1+K_{\mathrm{I}}}} \mathbf{H}_{\mathrm{I}, \mathrm{L}}+\sqrt{\frac{A_{\mathrm{N}} d_{\mathrm{I}}^{-\alpha_{\mathrm{N}}}}{1+K_{\mathrm{I}}}} \mathbf{H}_{\mathrm{I}, \mathrm{N}},
\end{aligned}
\end{equation}
where $K_{\text{U}}$, $K_{\text{E}}$ and $K_{\text{I}}$ are the Rician K-factors of the UBS-Alice channel, UBS-Eve channel and UBS-IRS channel, respectively. $A_{\text{L}}$ and $A_{\text{N}}$ are the path loss factors of the LOS and NLOS, respectively, $\alpha_{\mathrm{L}}$ and $\alpha_{\mathrm{N}}$ are the path loss exponents of the LOS and NLOS, respectively. The distance between the UBS-Alice, UBS-Eve and UBS-IRS, e.i., $d_{\mathrm{U}}$, $d_{\mathrm{E}}$ and $d_{\mathrm{I}}$ are expressed as
\begin{equation}
\begin{aligned}
d_{\text{U}}=\sqrt{\left(z_{\text{U}}\right)^{2}+\left(y_{\text{U}}-y_{\text{A}}\right)^{2} + \left(x_{\text{U}}-x_{\text{A}}\right)^{2} } , \\
d_{\text{E}}=\sqrt{\left(z_{\text{U}}\right)^{2}+\left(y_{\text{U}}-y_{\text{E}}\right)^{2}+\left(x_{\text{U}}-x_{\text{E}}\right)^{2} } , \\
d_{\text{I}}=\sqrt{\left(z_{\text{U}}-z_{\text{I}}\right)^{2}+\left(y_{\text{U}}-y_{\text{I}}\right)^{2}+\left(x_{\text{U}}-x_{\text{I}}\right)^{2}  }. \\
\end{aligned}
\end{equation}
In addition, the channel fading between the IRS and Alice/Eve also follows Rician distribution which can be generated with a similar procedure, and denoted as $\mathbf{h}_{\mathrm{IU}}$ and $\mathbf{h}_{\mathrm{IE}}$, respectively. Furthermore, in \eqref{channel}, the LOS and NLOS components of UBS-Alice channel $\mathbf{h}_{\mathrm{U}}$ are denoted as $\mathbf{h}_{\mathrm{U}, \mathrm{L}}$ and $\mathbf{h}_{\mathrm{U}, \mathrm{N}}$, respectively, and the UBS-Eve and UBS-IRS channels are also represented in a similar way. The NLOS component $\mathbf{h}_{\mathrm{U}, \mathrm{N}}$ with the variables independently is drawn from the circularly symmetric complex Gaussian distribution with zero mean and unit variance, i.e., $\mathcal{C} \mathcal{N}(0,1)$. The deterministic LOS component $\mathbf{h}_{\mathrm{U}, \mathrm{L}}$ and $\mathbf{h}_{\mathrm{U}, \mathrm{L}}$ are expressed as
\begin{equation}
\begin{aligned}
\label{h_U_E}
\mathbf{h}_{\mathrm{U}, \mathrm{L}}\!=&\left[1, e^{-2 \pi j \frac{b_{\mathrm{B}}}{\lambda}\left(\cos \omega_{\mathrm{U}}+\sin \phi_{\mathrm{U}} \sin \omega_{\mathrm{U}}\right)}, \cdots,\right.\\ &\left. e^{-2 \pi j \frac{b_{\mathrm{B}}}{\lambda}\left(N_{x}-1\right) \cos \omega_{\mathrm{U}}+\left(N_{y}-1\right) \sin \phi_{\mathrm{U}} \sin \omega_{\mathrm{U}}}\right]^{T}, \\
\mathbf{h}_{\mathrm{E}, \mathrm{L}}\!=&\left[1, e^{-2 \pi j \frac{b_{\mathrm{B}}}{\lambda}\left(\cos \omega_{\mathrm{E}}+\sin \phi_{\mathrm{E}} \sin \omega_{\mathrm{E}}\right)}, \cdots, \right.\\ &\left. e^{-2 \pi j \frac{b_{\mathrm{B}}}{\lambda}\left(N_{x}-1\right) \cos \omega_{\mathrm{E}}+\left(N_{y}-1\right) \sin \phi_{\mathrm{E}} \sin \omega_{\mathrm{E}}}\right]^{T},
\end{aligned}
\end{equation}
where $\omega_{\mathrm{U}} (\omega_{\mathrm{E}})$ and $\phi_{\mathrm{U}} (\phi_{\mathrm{E}})$ are the azimuth and elevation AOD of the path between the URA at UBS and Alice (Eve), respectively, $b_{\mathrm{B}}$ is the distance between two adjacent UBS antennas, and $\lambda$ is the wavelength of the center frequency of the carrier. Similarly, $\mathbf{H}_{\mathrm{I}, \mathrm{L}}$ is given by $\mathbf{H}_{\mathrm{I}, \mathrm{L}}=\mathbf{h}_{\mathrm{I}, \mathrm{L}}^{(\text{A})} \mathbf{h}_{\mathrm{I}, \mathrm{L}}^{(\text{D})}$, where
\begin{equation}
\begin{aligned}
&\mathbf{h}_{\mathrm{I}, \mathrm{L}}^{(\text{A})}\!=\!\left[1, e^{-2 \pi j \frac{b_{\mathrm{IRS}}}{\lambda}\left(\cos \varphi_{\mathrm{I}} \sin \vartheta_{\mathrm{I}}\!+\sin \varphi_{\mathrm{I}} \sin \vartheta_{\mathrm{I}}\right)},\right.\\ &\left. \!\cdots,\! e^{-2 \pi j \frac{b_{\mathrm{IRS}}}{\lambda}\left(M_{x}-1\right) \cos \varphi_{\mathrm{I}} \sin \vartheta_{\mathrm{I}}\!+\left(M_{y}-1\right) \sin \varphi_{\mathrm{I}} \sin \vartheta_{\mathrm{I}}}\right]^{T}
\end{aligned}
\end{equation}
is the array response at IRS with $\varphi_{\mathrm{I}}$ and $\vartheta_{\mathrm{I}}$ defined as the azimuth and elevation angle-of-arrival elevation (AOA) of the path between the URA at UBS and IRS. $b_{\mathrm{IRS}}$ is the antenna spacing of the URA at IRS.
\begin{equation}
\begin{aligned}
\label{h_I}
\mathbf{h}_{\mathrm{I}, \mathrm{L}}^{(\text{D})}\!=&\!\left[1, \!\cdots,\! e^{-2 \pi \frac{b_{\mathrm{B}}}{\lambda}\left(\cos \omega_{\mathrm{I}}\!+\sin \phi_{\mathrm{I}} \sin \omega_{\mathrm{I}}\right)},\!\cdots,\! \right.\\ &\left. e^{-2 \pi j \frac{b_{\mathrm{B}}}{\lambda}\left(N_{x}-1\right) \cos \omega_{\mathrm{I}}+\left(N_{y}-1\right) \sin \phi_{\mathrm{I}} \sin \omega_{\mathrm{I}}}\right]^{T}
\end{aligned}
\end{equation}
is the array response at UBS with $\omega_{\mathrm{I}}$ and $\phi_{\mathrm{I}}$ defined as the azimuth and elevation AOD of the path between the URA at UBS and IRS, respectively.
\subsection{CSI Error Models Considering UAV Jittering}
There is no doubt that the vibration characteristics of UAV are an important factor affecting the stability of its secrecy communication performance. In order to fully evaluate the impact of UAV jittering, we will establish a bounded CSI error model considering the UAV jittering in this section.

\begin{figure*}[tp]
\centering
\subfigure[] %子图片标题
{\includegraphics[width=2.3in]{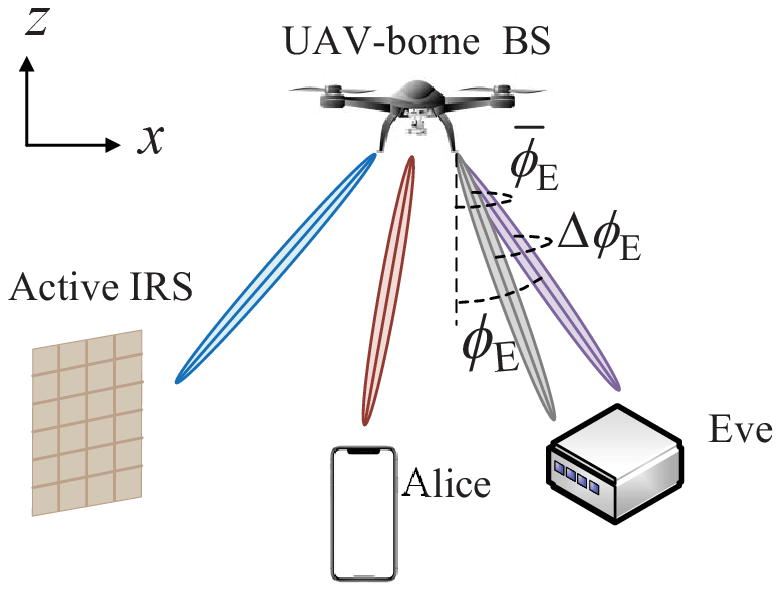}} %[图片大小]{图片路径}
\subfigure[]
{\includegraphics[width=2.4in]{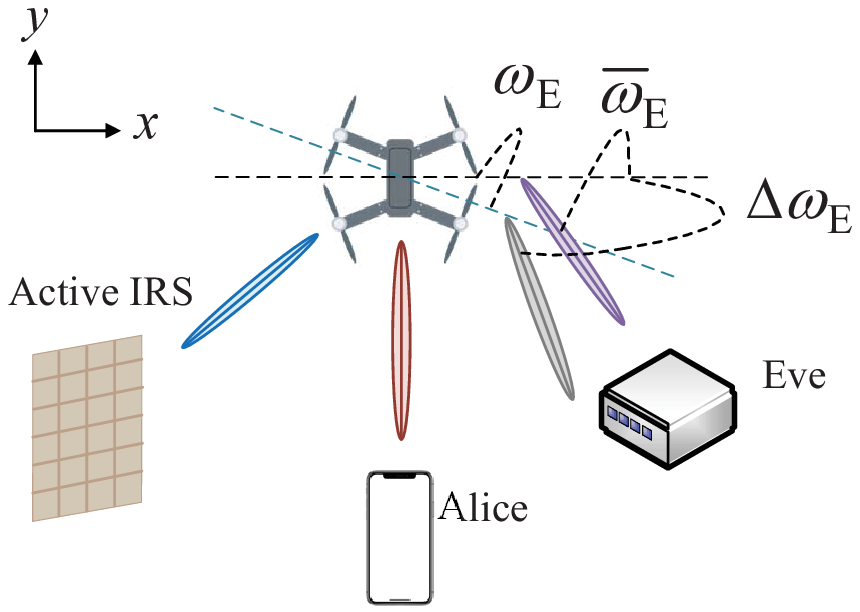}}
\caption{(a) Jittering of elevation angle. (b) Jittering of azimuth angle.} %图片标题
\label{system_CSI}  %图片交叉引用时的标签
\end{figure*}

The vibration characteristics of UAVs are caused by multiple factors. In addition to the torsional vibration excitation generated during the rotation of the engine-driven transmission system, the rotor-winged is also one of the vibration excitations. That means the UAV itself is a vibration system with free vibration and forced vibration [3]. Moreover, due to the randomness of wind gusts, rotation, yaw, and pitch motions, the stability of the UAV platform cannot be guaranteed. These random vibration characteristics will lead to imperfect CSI estimation and unstable wireless transmission, especially when equipped with a large antenna array. As shown in the Fig. \ref{system_CSI}(b) and Fig. \ref{system_CSI}(c), for a URA, varying elevation angles capture the UAV jittering in pitch angle and roll angle, while varying azimuth angles capture the UAV jittering in yaw angle.

To capture the jittering effect, the actual elevation AOD $\phi_{\mathrm{U}}$, $\phi_{\mathrm{E}}$ and $\phi_{\mathrm{I}}$, and the azimuth angle AOD $\omega_{\mathrm{U}}$ , $\omega_{\mathrm{E}}$ and $\omega_{\mathrm{I}}$ are modeled as
\begin{equation}
\label{angle_CSI}
\begin{aligned}
&\omega_{\mathrm{U}}=\bar{\omega}_{\mathrm{U}}+\Delta \omega_{\mathrm{U}}, \phi_{\mathrm{U}}=\bar{\phi}_{\mathrm{U}}+\Delta \phi_{\mathrm{U}}. \\
&\Omega_{\mathrm{U}}=\left\{\Delta \omega_{\mathrm{U}}, \Delta \phi_{\mathrm{U}} \in \mathbb{R} \mid\left(\Delta \omega_{\mathrm{U}}\right)^{2} \leqslant \beta_{\mathrm{U} 1}^{2},\left(\Delta \phi_{\mathrm{U}}\right)^{2} \leqslant \beta_{\mathrm{U} 2}^{2}\right\}. \\
&\omega_{\mathrm{E}}=\bar{\omega}_{\mathrm{E}}+\Delta \omega_{\mathrm{E}}, \phi_{\mathrm{E}}=\bar{\phi}_{\mathrm{E}}+\Delta \phi_{\mathrm{E}}. \\
&\Omega_{\mathrm{E}}=\left\{\Delta \omega_{\mathrm{E}}, \Delta \phi_{\mathrm{E}} \in \mathbb{R} \mid\left(\Delta \omega_{\mathrm{E}}\right)^{2} \leqslant \beta_{\mathrm{E} 1}^{2},\left(\Delta \phi_{\mathrm{E}}\right)^{2} \leqslant \beta_{\mathrm{E} 2}^{2}\right\}.\\
&\omega_{\mathrm{I}}=\bar{\omega}_{\mathrm{I}}+\Delta \omega_{\mathrm{I}}, \phi_{\mathrm{I}}=\bar{\phi}_{\mathrm{I}}+\Delta \phi_{\mathrm{I}}. \\
&\Omega_{\mathrm{I}}=\left\{\Delta \omega_{\mathrm{I}}, \Delta \phi_{\mathrm{I}} \in \mathbb{R} \mid\left(\Delta \omega_{\mathrm{I}}\right)^{2} \leqslant \beta_{\mathrm{I} 1}^{2},\left(\Delta \phi_{\mathrm{I}}\right)^{2} \leqslant \beta_{\mathrm{I} 2}^{2}\right\}. \\
\end{aligned}
\end{equation}
where $\bar{\omega}_{\mathrm{U}} (\bar{\omega}_{\mathrm{E}}, \bar{\omega}_{\mathrm{I}})$ and $\Delta \omega_{\mathrm{U}} (\Delta \omega_{\mathrm{E}}, \Delta \omega_{\mathrm{I}})$ are the estimated azimuth AOD and the uncertainty of azimuth AOD of the path between the UBS and Alice (Eve, IRS), respectively. $\bar{\phi}_{\mathrm{U}} (\bar{\phi}_{\mathrm{E}}, \bar{\phi}_{\mathrm{I}})$ and $\Delta \phi_{\mathrm{U}} (\Delta \phi_{\mathrm{E}}, \Delta \phi_{\mathrm{I}})$ are the estimated elevation AOD and the uncertainty of elevation AOD of the path between the UBS and Alice (Eve, IRS), respectively. $\Omega_{\mathrm{U}}$, $\Omega_{\mathrm{E}}$ and $\Omega_{\mathrm{B}}$ are the sets containing all the possible AOD uncertainties of the Alice, EVE and IRS, respectively.
The uncertainties of azimuth AOD and elevation AOD of Alice are bounded by the maximum variation $\beta_{\mathrm{U} 1}^{2}$ and $\beta_{\mathrm{U} 2}^{2}$, respectively, and the uncertainties of of Eve and IRS are represented in a similar way.
We note that $\mathbf{h}_{\mathrm{U}, \mathrm{L}} (\mathbf{h}_{\mathrm{E}, \mathrm{L}}, \mathbf{h}_{\mathrm{I}, \mathrm{L}})$ is a nonlinear function with respect to $\Delta \omega_{\mathrm{U}} (\Delta \omega_{\mathrm{E}}, \Delta \omega_{\mathrm{I}})$ and $\Delta \phi_{\mathrm{U}} (\Delta \phi_{\mathrm{E}}, \Delta \phi_{\mathrm{I}})$. To tackle this problem, we approximate $\mathbf{h}_{\mathrm{U}, \mathrm{L}} (\mathbf{h}_{\mathrm{E}, \mathrm{L}}, \mathbf{h}_{\mathrm{I}, \mathrm{L}}^{(\text{D})})$ by applying Taylor expansion, i.e.,
\begin{equation}
f(x+\Delta x, y+\Delta y) \approx f(x, y)+\Delta x \frac{\partial f(x, y)}{\partial x}+\Delta y \frac{\partial f(x, y)}{\partial y}.
\end{equation}
Thus, each exponential term in (\ref{h_U_E}) and (\ref{h_I}) can be approximated as (\ref{exp}).
\newcounter{mytempeqncnt}  %图片占双栏置顶
\begin{figure*}[!t]
\normalsize
\begin{equation}
\begin{aligned}
&\exp \left(-2 \pi j \frac{b_{\mathrm{B}}}{\lambda}\left(n_{x} \cos (\bar{\omega}+\Delta \omega)+n_{y} \sin (\bar{\phi}+\Delta \phi) \sin (\bar{\omega}+\Delta \omega)\right)\right) \\
&\approx \exp \left(-2 \pi j \frac{b_{\mathrm{B}}}{\lambda}\left(n_{x} \cos (\omega)+n_{y} \sin (\bar{\phi}) \sin (\bar{\omega})\right)+\exp \left(-2 \pi j \frac{b_{\mathrm{B}}}{\lambda}\left(n_{x} \cos (\omega)+n_{y} \sin (\bar{\phi}) \sin (\bar{\omega})\right)\right)\right. \\
&\times\left(2 \pi j \frac{b_{\mathrm{B}}}{\lambda}\left(n_{x} \sin (\bar{\omega})-n_{y} \sin (\bar{\phi}) \cos (\bar{\omega})\right) \Delta \omega-2 \pi j \frac{b_{\mathrm{B}}}{\lambda} n_{x} \cos (\bar{\phi}) \sin (\bar{\omega}) \Delta \phi\right).
\label{exp}
\end{aligned}
\end{equation}
\hrulefill
\vspace*{4pt}
\end{figure*}

We define
\begin{equation}
\begin{aligned}
\mathbf{a}(\bar{\phi}, \bar{\omega})\!&=\!\sin (\bar{\omega})\left[0,2 \pi j \frac{b_{\mathrm{B}}}{\lambda}, \ldots, 2 \pi j \frac{b_{\mathrm{B}}}{\lambda}\left(N_{x}-1\right)\right] \\
&-\sin (\bar{\phi}) \cos (\bar{\omega})\left[0,2 \pi j \frac{b_{\mathrm{B}}}{\lambda}, \ldots, 2 \pi j \frac{b_{\mathrm{B}}}{\lambda}\left(N_{y}-1\right)\right], \\
\end{aligned}
\end{equation}
\begin{equation}
\begin{aligned}
\mathbf{b}(\bar{\phi}, \bar{\omega})\!=\!\cos (\bar{\phi})\sin (\bar{\omega})\left[0,-2 \pi j \frac{b_{\mathrm{B}}}{\lambda}, \ldots, -2 \pi j \frac{b_{\mathrm{B}}}{\lambda}\left(N_{y}\!-\!1\right)\right].
\end{aligned}
\end{equation}
Then,  $\mathbf{h}_{\mathrm{U}, \mathrm{L}}, \mathbf{h}_{\mathrm{E}, \mathrm{L}}$ and $\mathbf{h}_{\mathrm{I}, \mathrm{L}}^{(\text{D})}$ in Eq. (3) and Eq. (5) can be given as
\begin{equation}
\begin{aligned}
\mathbf{h}_{\mathrm{U}, \mathrm{L}} &= \overline{\mathbf{h}}_{\mathrm{U}, \mathrm{L}}+ \Delta {\mathbf{h}}_{\mathrm{U}, \mathrm{L}}, \\
\mathbf{h}_{\mathrm{E}, \mathrm{L}} &= \overline{\mathbf{h}}_{\mathrm{E}, \mathrm{L}}+ \Delta {\mathbf{h}}_{\mathrm{E}, \mathrm{L}}, \\
\mathbf{h}_{\mathrm{I}, \mathrm{L}}^{(\mathrm{D})} &= \overline{\mathbf{h}}_{\mathrm{I}, \mathrm{L}}^{(\mathrm{D})}+ \Delta {\mathbf{h}}_{\mathrm{I}, \mathrm{L}}^{(\mathrm{D})},
\end{aligned}
\end{equation}where
\begin{equation}
\begin{aligned}
\Delta {\mathbf{h}}_{\mathrm{U}, \mathrm{L}} &\approx  \overline{\mathbf{h}}_{\mathrm{U}, \mathrm{L}}\odot \mathbf{a}\left(\bar{\phi}_{\mathrm{U}}, \bar{\omega}_{\mathrm{U}}\right) \Delta \omega_{\mathrm{U}} +\overline{\mathbf{h}}_{\mathrm{U}, \mathrm{L}} \odot \mathbf{b}\left(\bar{\phi}_{\mathrm{U}}, \bar{\omega}_{\mathrm{U}}\right) \Delta \phi_{\mathrm{U}}, \\
\Delta {\mathbf{h}}_{\mathrm{E}, \mathrm{L}}  &\approx \overline{\mathbf{h}}_{\mathrm{E}, \mathrm{L}} \odot \mathbf{a}\left(\bar{\phi}_{\mathrm{E}}, \bar{\omega}_{\mathrm{E}}\right) \Delta \omega_{\mathrm{E}} +\overline{\mathbf{h}}_{\mathrm{E}, \mathrm{L}} \odot \mathbf{b}\left(\bar{\phi}_{\mathrm{E}}, \bar{\omega}_{\mathrm{E}}\right) \Delta \phi_{\mathrm{E}}, \\
\Delta {\mathbf{h}}_{\mathrm{I}, \mathrm{L}}^{(\mathrm{D})}  & \approx \overline{\mathbf{h}}_{\mathrm{I}, \mathrm{L}}^{(\mathrm{D})} \odot \mathbf{a}\left(\bar{\phi}_{\mathrm{I}}, \bar{\omega}_{\mathrm{I}}\right) \Delta \omega_{\mathrm{I}} +\overline{\mathbf{h}}_{\mathrm{I}, \mathrm{L}}^{(\mathrm{D})} \odot \mathbf{b}\left(\bar{\phi}_{\mathrm{I}}, \bar{\omega}_{\mathrm{I}}\right) \Delta \phi_{\mathrm{I}},
\end{aligned}
\end{equation}
 and $\overline{\mathbf{h}}_{\mathrm{U}, \mathrm{L}}, \overline{\mathbf{h}}_{\mathrm{E}, \mathrm{L}}$ and $\overline{\mathbf{h}}_{\mathrm{I}, \mathrm{L}}^{(\mathrm{D})}$ are the estimated channel vectors represented by
\begin{equation}
\begin{aligned}
&\overline{\mathbf{h}}_{\mathrm{U}, \mathrm{L}}=\left[0, e^{-2 \pi j \frac{b_{\mathrm{B}}}{\lambda}\left(\cos \bar{\omega}_{\mathrm{U}}+\sin \bar{\phi}_{\mathrm{U}}\sin \bar{\omega}_{\mathrm{U}}\right)}, \cdots, \right. \\
&\left.e^{-2 \pi j \frac{b_{\mathrm{B}}}{\lambda}\left(\left(N_{x}-1\right) \cos \bar{\omega}_{\mathrm{U}}+\left(N_{y}-1\right) \sin \bar{\phi}_{\mathrm{U}} \sin \bar{\omega}_{\mathrm{U}}\right)}\right], \\
&\overline{\mathbf{h}}_{\mathrm{E}, \mathrm{L}}=\left[0, e^{-2 \pi j \frac{b_{\mathrm{B}}}{\lambda}\left(\cos\bar{\omega}_{\mathrm{E}}+\sin \bar{\phi}_{\mathrm{E}} \sin \bar{\omega}_{\mathrm{E}}\right)}, \cdots, \right. \\
&\left.e^{-2 \pi j \frac{b_{\mathrm{B}}}{\lambda}\left(\left(N_{x}-1\right) \cos \bar{\omega}_{\mathrm{E}}+\left(N_{y}-1\right) \sin \bar{\phi}_{\mathrm{E}} \sin \bar{\omega}_{\mathrm{E}}\right)}\right], \\
&\overline{\mathbf{h}}_{\mathrm{I}, \mathrm{L}}^{(\mathrm{D})}=\left[0, e^{-2 \pi j \frac{b_{\mathrm{B}}}{\lambda}\left(\cos \bar{\omega}_{\mathrm{I}}+\sin \bar{\phi}_{\mathrm{I}} \sin \bar{\omega}_{\mathrm{I}}\right)}, \cdots, \right. \\
&\left.e^{-2 \pi j \frac{b_{\mathrm{B}}}{\lambda}\left(\left(N_{x}-1\right) \cos \bar{\omega}_{\mathrm{I}}+\left(N_{y}-1\right) \sin \bar{\phi}_{\mathrm{I}} \sin \bar{\omega}_{\mathrm{I}}\right)}\right],
\end{aligned}
\end{equation}
where
\begin{equation}
\begin{aligned}
&\sin \bar{\phi}_{\text{U}}\sin \bar{\omega}_{\text{U}}=\frac{y_{\text{U}}-y_{\text{A}}}{d_{\text{U}}},\cos\bar{\omega}_{\text{U}}=\frac{z_{\text{U}}-z_{\text{A}}}{d_{\text{U}}}, \\
&\sin \bar{\phi}_{\text{E}} \sin \bar{\omega}_{\text{E}}=\frac{y_{\text{U}}-y_{\text{E}}}{d_{\text{E}}},\cos\bar{\omega}_{\text{E}}=\frac{z_{\text{U}}-z_{\text{E}}}{d_{\text{E}}}, \\
 &\sin \bar{\phi}_{\text{I}} \sin \bar{\omega}_{\text{I}}=\frac{y_{\text{U}}-y_{\text{I}}}{d_{\text{I}}},\cos\bar{\omega}_{\text{I}}=\frac{z_{\text{U}}-z_{\text{I}}}{d_{\text{I}}}, \\
&\sin \varphi_{\mathrm{B}} \sin \vartheta_{\mathrm{B}}\!=\! \frac{y_{\text{U}}-y_{\text{I}}}{d_{\text{I}}},\cos \varphi_{\mathrm{B}} \sin \vartheta_{\mathrm{B}} \!=\! \frac{x_{\text{U}}-x_{\text{I}}}{d_{\text{I}}}.
\end{aligned}
\end{equation}
Furthermore, the channel variations caused by UAV jittering $\Delta \mathbf{h}_{\mathrm{U}, \mathrm{L}}, \Delta \mathbf{h}_{\mathrm{E}, \mathrm{L}}$ and $\Delta \mathbf{h}_{\mathrm{I}, \mathrm{L}}^{(\mathrm{D})}$ are given by
\begin{equation}
\begin{aligned}
\notag\Delta \mathbf{h}_{\mathrm{U}} &=\sqrt{\frac{A_{\mathrm{L}} d_{\mathrm{U}}^{-\alpha_{\mathrm{L}}} K_{\mathrm{U}}}{1+K_{\mathrm{U}}}} \Delta \mathbf{h}_{\mathrm{U}, \mathrm{L}}  \triangleq \mathbf{a}_{\mathrm{U}} \Delta \omega_{\mathrm{U}}+\mathbf{b}_{\mathrm{U}} \Delta \phi_{\mathrm{U}}, \\
\Delta \mathbf{h}_{\mathrm{E}} &=\sqrt{\frac{A_{\mathrm{L}} d_{\mathrm{E}}^{-\alpha_{\mathrm{L}}} K_{\mathrm{E}}}{1+K_{\mathrm{E}}}} \Delta \mathbf{h}_{\mathrm{E}, \mathrm{L}} \triangleq \mathbf{a}_{\mathrm{E}} \Delta \omega_{\mathrm{E}}+\mathbf{b}_{\mathrm{E}} \Delta \phi_{\mathrm{E}},
\end{aligned}
\end{equation}
\begin{equation}
\begin{aligned}
\Delta \mathbf{H}_{\mathrm{I}} &=\mathbf{h}_{\mathrm{I}, \mathrm{L}}^{(\mathrm{A})} \sqrt{\frac{A_{\mathrm{L}} d_{\mathrm{I}}^{-\alpha_{\mathrm{L}}} K_{\mathrm{I}}}{1+K_{\mathrm{I}}}} \Delta \mathbf{h}_{\mathrm{I}, \mathrm{L}}^{(\mathrm{D})} \triangleq \mathbf{h}_{\mathrm{I}, \mathrm{L}}^{(\mathrm{A})}\left(\mathbf{a}_{\mathrm{I}} \Delta \omega_{\mathrm{I}}+\mathbf{b}_{\mathrm{I}} \Delta \phi_{\mathrm{I}}\right).
\end{aligned}
\end{equation}
Thus, $\mathbf{h}_{\mathrm{U}}, \mathbf{h}_{\mathrm{E}}$ and $\mathbf{H}_{\mathrm{I}}$ in Eq. (1) and Eq. (2) can be rewritten as
\begin{equation}
\begin{aligned}
\mathbf{h}_{\mathrm{U}}&=\underbrace{\sqrt{\frac{A_{\mathrm{L}} d_{\mathrm{U}}^{-\alpha_{\mathrm{L}}} K_{\mathrm{U}}}{1+K_{\mathrm{U}}}} \overline{\mathbf{h}}_{\mathrm{U}, \mathrm{L}}+\sqrt{\frac{A_{\mathrm{N}} d_{\mathrm{U}}^{-\alpha_{\mathrm{N}}}}{1+K_{\mathrm{U}}}} \mathbf{h}_{\mathrm{U}, \mathrm{N}}}_{\overline{\mathbf{h}}_{\mathrm{U}}}+\Delta \mathbf{h}_{\mathrm{U}},\\
\mathbf{h}_{\mathrm{E}}&=\underbrace{\sqrt{\frac{A_{\mathrm{L}} d_{\mathrm{E}}^{-\alpha_{\mathrm{L}}} K_{\mathrm{E}}}{1+K_{\mathrm{E}}}} \overline{\mathbf{h}}_{\mathrm{E}, \mathrm{L}}+\sqrt{\frac{A_{\mathrm{N}} d_{\mathrm{E}}^{-\alpha_{\mathrm{N}}}}{1+K_{\mathrm{E}}}} \mathbf{h}_{\mathrm{E}, \mathrm{N}}}_{\overline{\mathbf{h}}_{\mathrm{E}}}+\Delta \mathbf{h}_{\mathrm{E}},\\
\mathbf{H}_{\mathrm{I}}&= \underbrace{\sqrt{\frac{A_{\mathrm{L}} d_{\mathrm{I}}^{-\alpha_{\mathrm{L}}} K_{\mathrm{I}}}{1+K_{\mathrm{I}}}} \mathbf{h}_{\mathrm{I}, \mathrm{L}}^{(\mathrm{A})} \overline{\mathbf{h}}_{\mathrm{I}, \mathrm{L}}^{(\mathrm{D})}+\sqrt{\frac{A_{\mathrm{N}} d_{\mathrm{I}}^{-\alpha_{\mathrm{N}}}}{1+K_{\mathrm{I}}}} \mathbf{H}_{\mathrm{I}, \mathrm{N}}}_{\overline{\mathbf{H}}_{\mathrm{I}}}+\Delta \mathbf{H}_{\mathrm{I}}.
\end{aligned}
\end{equation}

In order to evaluate the bounded uncertainty of the channel caused by UAV jittering, we further derive the CSI error model of $\mathbf{h}_{\mathrm{U}}$ and $\mathbf{h}_{\mathrm{E}}$, which is given as
\begin{equation}
\label{channel_CSI}
\begin{aligned}
\left\|\Delta \mathbf{h}_{\mathrm{U}}\right\|_{2}&\!=\!\left\|\mathbf{a}_{\mathrm{U}}\Delta \omega_{\mathrm{U}}\!+\!\mathbf{b}_{\mathrm{U}} \Delta \phi_{\mathrm{U}}\right\|_{2} \!\leq\!\left\|\mathbf{a}_{\mathrm{U}} \Delta \omega_{\mathrm{U}}\right\|_{2}\!+\!\left\|\mathbf{b}_{\mathrm{U}} \Delta \phi_{\mathrm{U}}\right\|_{2}, \\
\left\|\Delta \mathbf{h}_{\mathrm{E}}\right\|_{2}&\!=\!\left\|\mathbf{a}_{\mathrm{E}}\Delta \omega_{\mathrm{E}}\!+\!\mathbf{b}_{\mathrm{E}} \Delta \phi_{\mathrm{E}}\right\|_{2}\!\leq\!\left\|\mathbf{a}_{\mathrm{E}} \Delta \omega_{\mathrm{E}}\right\|_{2}\!+\!\left\|\mathbf{b}_{\mathrm{E}} \Delta \phi_{\mathrm{E}}\right\|_{2}.
\end{aligned}
\end{equation}
To further simplify the expression, by means of variable substitution $\left\|\mathbf{a}_{\mathrm{U}}\right\|_{2}=a_{\mathrm{U} 1},\left\|\mathbf{b}_{\mathrm{U}}\right\|_{2}=a_{\mathrm{U} 2},\left\|\mathbf{a}_{\mathrm{E}}\right\|_{2}=a_{\mathrm{E} 1},\left\|\mathbf{b}_{\mathrm{E}}\right\|_{2}=a_{\mathrm{E} 2}$
%$\left\|\mathbf{H}_{\mathrm{I},\mathrm{L}}^{(\mathrm{A})}\right\|_{F}\left\|\mathbf{a}_{\mathrm{B}}\right\|_{2}=a_{\mathrm{B} 1}$,
%$\left\|\mathbf{H}_{\mathrm{I}, \mathrm{L}}^{(\mathrm{A})}\right\|_{F}\left\|\mathbf{b}_{\mathrm{B}}\right\|_{2}=a_{\mathrm{B} 2}$
 and \eqref{angle_CSI}, \eqref{channel_CSI} can be rewritten as
\begin{equation}
\begin{aligned}
\label{H_CSI}
&\left\|\Delta \mathbf{h}_{\mathrm{U}}\right\|_{2} \leq\beta_{\mathrm{U} 1} a_{\mathrm{U} 1}+\beta_{\mathrm{U} 2} a_{\mathrm{U} 2} \triangleq \xi_{\mathrm{U},\mathrm{h}}, \\
&\left\|\Delta \mathbf{h}_{\mathrm{E}}\right\|_{2} \leq\beta_{\mathrm{E} 1} a_{\mathrm{E} 1}+\beta_{\mathrm{E} 2} a_{\mathrm{E} 2} \triangleq \xi_{\mathrm{E},\mathrm{h}}. \\
%&\left\|\Delta \mathbf{H}_{\mathrm{I}}\right\|_{F} \leq\beta_{\mathrm{B} 1} a_{\mathrm{B} 1}+\beta_{\mathrm{B} 2} a_{\mathrm{B} 2}.
\end{aligned}
\end{equation}
Similarly, the CSI error model of $\mathbf{H}_{\mathrm{I}}$ is given as
\begin{equation}
\begin{aligned}
\left\|\Delta \mathbf{H}_{\mathrm{I}}\right\|_{F}&=\left\|\mathbf{H}_{\mathrm{I}, \mathrm{L}}^{(\mathrm{A})}\left(\mathbf{a}_{\mathrm{B}} \Delta \omega_{\mathrm{B}}+\mathbf{b}_{\mathrm{B}} \Delta \phi_{\mathrm{B}}\right)\right\|_{F},\\
&\leq\left|\Delta \omega_{\mathrm{B}}\right|\left\|\mathbf{H}_{\mathrm{I}, \mathrm{L}}^{(\mathrm{A})}\right\|_{F}\left\|\mathbf{a}_{\mathrm{B}}\right\|_{2}\!+\!\left|\Delta \phi_{\mathrm{B}}\right|\left\|\mathbf{H}_{\mathrm{I}, \mathrm{L}}^{(\mathrm{A})}\right\|_{F}\left\|\mathbf{b}_{\mathrm{B}}\right\|_{2}.
\end{aligned}
\end{equation}
In order to facilitate subsequent analysis, the channel UBS-IRS-Alice and UBS-IRS-Eve are represented by the cascaded channel $\mathbf{G}_{\mathrm{U}}$ and $\mathbf{G}_{\mathrm{E}}$, respectively. Denote by $\mathbf{G}_{\mathrm{U}}=\operatorname{diag}\left(\mathbf{h}_{\mathrm{IU}}^{H}\right) \mathbf{H}_{\mathrm{I}}$ and $\mathbf{G}_{\mathrm{E}}=\operatorname{diag}\left(\mathbf{h}_{\mathrm{IE}}^{H}\right) \mathbf{H}_{\mathrm{I}}$ the cascaded channel via IRS. The CSI error model of the cascaded channel $\mathbf{G}_{\mathrm{U}}$ and $\mathbf{G}_{\mathrm{E}}$ is expressed as
\begin{align}
\label{G_CSI}
\notag \left\|\Delta \mathbf{G}_{\mathrm{U}}\right\|_{F} \leq\beta_{\mathrm{B} 1} g_{\mathrm{U} 1}+\beta_{\mathrm{B} 2}g_{\mathrm{U} 2} \triangleq \xi_{\mathrm{U},\mathrm{G}}, \\
\left\|\Delta \mathbf{G}_{\mathrm{E}}\right\|_{F} \leq\beta_{\mathrm{B} 1} g_{\mathrm{E} 1}+\beta_{\mathrm{B} 2} g_{\mathrm{E} 2} \triangleq \xi_{\mathrm{E},\mathrm{G}}.
\end{align}
where
\begin{equation}
\begin{aligned}
g_{\mathrm{U} 1}=\left\|\operatorname{diag}\left(\mathbf{h}_{\mathrm{IU}}^{H}\right)\right\|_{F} \left\|\mathbf{H}_{\mathrm{I},\mathrm{L}}^{(\mathrm{A})}\right\|_{F}\left\|\mathbf{a}_{\mathrm{B}}\right\|_{2}, \\
g_{\mathrm{U} 2}=\left\|\operatorname{diag}\left(\mathbf{h}_{\mathrm{IU}}^{H}\right)\right\|_{F} \left\|\mathbf{H}_{\mathrm{I}, \mathrm{L}}^{(\mathrm{A})}\right\|_{F}\left\|\mathbf{b}_{\mathrm{B}}\right\|_{2}, \\
g_{\mathrm{E} 1}=\left\|\operatorname{diag}\left(\mathbf{h}_{\mathrm{IE}}^{H}\right)\right\|_{F} \left\|\mathbf{H}_{\mathrm{I},\mathrm{L}}^{(\mathrm{A})}\right\|_{F}\left\|\mathbf{a}_{\mathrm{B}}\right\|_{2}, \\
g_{\mathrm{E} 2}=\left\|\operatorname{diag}\left(\mathbf{h}_{\mathrm{IE}}^{H}\right)\right\|_{F} \left\|\mathbf{H}_{\mathrm{I},  \mathrm{L}}^{(\mathrm{A})}\right\|_{F}\left\|\mathbf{b}_{\mathrm{B}}\right\|_{2},
\end{aligned}
\end{equation}
\subsection{Signal Model}
The signal received at Alice from both the UBS-Alice and UBS-IRS-Alice channels is expressed as
\begin{equation}
\begin{aligned}
y_{\mathrm{U}} &=\left(\mathbf{h}_{\mathrm{IU}}^{H} \boldsymbol{\Theta} \mathbf{H}_{\mathrm{I}}+\mathbf{h}_{\mathrm{U}}^{H}\right) \mathbf{w} s+\mathbf{h}_{\mathrm{IU}}^{H} \boldsymbol{\Theta} \mathbf{n}_{\mathrm{I}}+n_{\mathrm{U}}, \\
\end{aligned}
\end{equation}
and the signal received at Eve is
\begin{equation}
\begin{aligned}
y_{\mathrm{E}} &=\left(\mathbf{h}_{\mathrm{IE}}^{H} \boldsymbol{\Theta} \mathbf{H}_{\mathrm{I}}+\mathbf{h}_{\mathrm{E}}^{H}\right) \mathbf{w} s+\mathbf{h}_{\mathrm{IE}}^{H} \boldsymbol{\Theta} \mathbf{n}_{\mathrm{I}}+n_{\mathrm{E}}, \\
\end{aligned}
\end{equation}
where $\mathbf{n}_{\mathrm{I}} \sim \mathcal{C} \mathcal{N}\left(0, \sigma_{\mathrm{I}}^{2} \mathbf{I}_{M}\right)$ denotes the additive white Gaussian noise (AWGN) at the IRS. $n_{\mathrm{U}} \sim \mathcal{C} \mathcal{N}\left(0, \sigma_{\mathrm{U}}^{2}\right), n_{\mathrm{E}}\sim \mathcal{C} \mathcal{N}\left(0, \sigma_{\mathrm{E}}^{2}\right)$ denotes the AWGN at the Alice and Eve, respectively. $s$ is the data symbol vector with unit-power and $\mathbf{w}$ is the transmit beamformer. Denote by $\mathbf{v}=\left[v_{1}, \ldots, v_{M}\right]^{T} \in \mathbb{C}^{M \times 1}$ the vector containing diagonal elements of matrix $\boldsymbol{\Theta}$ where $v_{m}=\tau_{m} e^{j \theta_{m}}, \forall m$. As such, the achievable rate of Alice and Eve are respectively given by
\begin{equation}
\begin{aligned}
R_{\mathrm{U}}=\log _{2}\left(1+\gamma_{\mathrm{U}}\right), \\
R_{\mathrm{E}}=\log _{2}\left(1+\gamma_{\mathrm{E}}\right),
\end{aligned}
\end{equation}
where
\begin{small}
\begin{equation}
\begin{aligned}
\gamma_{\mathrm{U}}&=\frac{\left|\left[\left(\overline{\mathbf{h}}_{\mathrm{U}}^{H}+\Delta \mathbf{h}_{\mathrm{U}}^{H}\right)+\mathbf{h}_{\mathrm{IU}}^{H} \boldsymbol{\Theta}\left(\overline{\mathbf{H}}_{\mathrm{I}}+\Delta \mathbf{H}_{\mathrm{I}}\right)\right] \mathbf{w}\right|^{2}}{\sigma_{\mathrm{I}}^{2}\left\|\mathbf{h}_{\mathrm{IU}}^{H} \boldsymbol{\Theta}\right\|^{2}+\sigma_{\mathrm{U}}^{2}} \\
&=\frac{\left|\left[\left(\overline{\mathbf{h}}_{\mathrm{U}}^{H}+\Delta \mathbf{h}_{\mathrm{U}}^{H}\right)+\mathbf{v}^{H}\left(\overline{\mathbf{G}}_{\mathrm{U}}+\Delta \mathbf{G}_{\mathrm{U}}\right)\right] \mathbf{w}\right|^{2}}{\sigma_{\mathrm{I}}^{2}\left\|\mathbf{v}^{H}\operatorname{diag}(\mathbf{h}_{\mathrm{IU}}^{H})\right\|^{2}+\sigma_{\mathrm{U}}^{2}}, \\
\gamma_{\mathrm{E}}&=\frac{\left|\left[\left(\overline{\mathbf{h}}_{\mathrm{E}}^{H}+\Delta \mathbf{h}_{\mathrm{E}}^{H}\right)+\mathbf{h}_{\mathrm{IE}}^{H} \boldsymbol{\Theta}\left(\overline{\mathbf{H}}_{\mathrm{I}}+\Delta \mathbf{H}_{\mathrm{I}}\right)\right] \mathbf{w}\right|^{2}}{\sigma_{\mathrm{I}}^{2}\left\|\mathbf{h}_{\mathrm{IE}}^{H} \boldsymbol{\Theta}\right\|^{2}+\sigma_{\mathrm{E}}^{2}} \\
&=\frac{\left|\left[\left(\overline{\mathbf{h}}_{\mathrm{E}}^{H}+\Delta \mathbf{h}_{\mathrm{E}}^{H}\right)+\mathbf{v}^{H}\left(\overline{\mathbf{G}}_{\mathrm{E}}+\Delta \mathbf{G}_{\mathrm{E}}\right)\right] \mathbf{w}\right|^{2}}{\sigma_{\mathrm{I}}^{2}\left\|\mathbf{v}^{H}\operatorname{diag}(\mathbf{h}_{\mathrm{IE}}^{H} )\right\|^{2}+\sigma_{\mathrm{E}}^{2}}.
\end{aligned}
\end{equation}
\end{small}

\section{Robust and Secure UAV Communication assisted by Active IRS}
In order to ensure ASR while improving UAV's power-efficiency, with the UAV jittering model, we first formulate the transmission power minimization problem under the worst-case secrecy rate constraints, UAV's maximum transmission power constraints, and the active IRS's amplification power constraint. Due to the coupling relationship of various optimization variables and complex non-convex constraints, the proposed problem is non-convex and difficult to solve. To solve this problem, we transform all the non-convex constraints into tractable LMIs forms by S-procedure and Schur's complement. Then the AO strategy is utilized to decouple the parameters and transform the original problems into two subproblems, i.e., passive beamforming and active IRS's reflecting coefficient optimization.
\subsection{Transmission Power Minimization Problem Formulation}
In practice, the transmission power of UAVs is strictly limited which is necessary to be reduce as much as possible while ensuring secure communication requirements. Thus, we focus on reducing the transmission power of the active IRS-assisted UAV system under the given ASR constraint, i.e., the worst-case scenario where Alice reaches its required minimum data rate and Eve reaches its maximum data rate. In order to ensure the secure transmission between the UBS and Alice, the worst-case secrecy rate of the considered system should be above 0, which is given by $R_{\mathrm{sec}}$
\begin{equation}
\begin{aligned}
R_{\mathrm{sec}}=\left[\min _{\left\{\Omega_{\mathrm{B}}, \Omega_{\mathrm{U}}\right\}} R_{\mathrm{U}}-\max _{\left\{\Omega_{\mathrm{B}}, \Omega_{\mathrm{E}}\right\}} R_{\mathrm{E}}\right]^{+}.
\end{aligned}
\end{equation}

 In this paper, we adopted an active IRS in which each reflecting element (RE) is supported by a set of active-load impedances. In this way, active IRS solves the double-fading attenuation problem not only by increasing the number of REs, but also by amplifying the incident signal at each RE. Due to the ability to amplify the incident signal, fewer number of REs is required to achieve the target signal-to-noise ratio. That means the physical size of the active IRS can be reduced, making it more suitable for UAV communication limited by the payload. Furthermore, it offers more flexibility to reconfigure the wireless propagation environment by optimizing the amplitudes of the reflection coefficients instead of just phases, resulting in a more spectrum and energy-efficient communication. Note that the active IRS amplifies both the received signal and noise at each RE. Besides, the amplification power is limited due to the overall power budget at the IRS \cite{Active_IRS}. Let $P_{\text{F}}$ denote the maximum amplification power of the active IRS, which is practically much smaller than the conventional RF amplifier due to the limited amplification power gain [11], [12]. Then we have
\begin{equation}
\label{F}
\left\|\boldsymbol{\Theta} \mathbf{H}_{\mathrm{I}} \mathbf{w}\right\|^{2}+\sigma_{\mathrm{I}}^{2}\left\|\boldsymbol{\Theta}\right\|^{2} \leq P_{\text{F}}.
\end{equation}

To minimize transmission power of UBS, the reflecting coefficient matrix $\boldsymbol{\Theta}$ of the active IRS and the transmit beamformer $\mathbf{w}$ are jointly optimized, and the corresponding optimization problem is formulated as follows.
\begin{subequations}
\label{P1}
\begin{align}
(\mathrm{P} 1): \notag \min_{\boldsymbol{\Theta}, \mathbf{w}} &\ \|\mathbf{w}\|^{2} \\
\label{C1}
\text { s.t. } &\|\mathbf{w}\|^{2}  \leq P_{\text {peak}}, \\
\label{C3}
& \min _{\left\{\Omega_{\mathrm{B}}, \Omega_{\mathrm{U}}\right\}} R_{\mathrm{U}} \geq \eta_{\mathrm{U}}, \\
\label{C4}
& \max _{\left\{\Omega_{\mathrm{B}}, \Omega_{\mathrm{E}}\right\}} R_{\mathrm{E}} \leq \eta_{\mathrm{E}}, \\
\label{C7}
&\left\|\boldsymbol{\Theta} \mathbf{H}_{\mathrm{I}} \mathbf{w}\right\|^{2}+\sigma_{\mathrm{I}}^{2}\left\|\boldsymbol{\Theta}\right\|^{2} \leq P_{\text{F}},\\
&|\boldsymbol{\Theta}[m, m]| \leq \tau_{\text{max}}
\end{align}
\end{subequations}
where $\mathbf{w}$ is constrained by a peak power limit $P_{\text {peak}}$ i.e., \eqref{C1}, and $\tau_{\text{max}}$ is the maximum amplification factor at the m-th reflecting element. Constraint \eqref{C3} ensures that the minimum data rate of the legitimate channel is above $\eta_{\mathrm{U}}(>0)$, i.e., ensures the effectiveness and reliability of the legitimate link. Constraint \eqref{C4} ensures that the maximum eavesdropping data rate is below $\eta_{\mathrm{E}}\left(0<\eta_{\mathrm{E}}<\eta_{\mathrm{U}}\right)$, i.e., ensures the security of the confidential message. \eqref{C3} and \eqref{C4} jointly guarantee the secure transmission requirement of the confidential signal. As mentioned before, \eqref{C7} is the amplification power constraint. It is observed that the \eqref{C3} and \eqref{C4} are the constraints for lower bound and upper bound, respectively. In addition, $\mathbf{w}$ and $\boldsymbol{\Theta}$ are coupled in \eqref{C7}. Thus, the optimization problem in \eqref{P1} is non-convex and challenging to solve. To cope with this difficulty, we propose an efficient algorithm based on the AO method to optimize $\mathbf{w}$ and $\boldsymbol{\Theta}$ sequentially in an iterative manner. Specifically, we divide problem P1 into two sub-problems:

1) The optimization of the transmit $\mathbf{w}$ under the given reflecting coefficient matrix $\boldsymbol{\Theta}$;

2) The optimization of the reflecting coefficient matrix $\boldsymbol{\Theta}$ under the given transmission power $\mathbf{w}$;
\subsection{Alternating Optimization Solution of the Problem}
In this section, we alternately solve the two sub-problems to minimize the transmission power at UBS. To deal with non-convex constraints, we first use S-procedure to convert the worst-case secrecy rate constraints into LMIs. Since the logarithmic function is a monotonically increasing function, by considering the channel uncertainty in \eqref{H_CSI} and \eqref{G_CSI} and denoting
 $\mathcal{E}^{\text {Alice }} \triangleq\left\{\forall\left\|\Delta \mathbf{h}_{\text U}\right\|_{2} \leq\right.$ $\left.\xi_{\mathrm{U},\mathrm{h}}, \forall\left\|\Delta \mathbf{G}_{\text U}\right\|_{F} \leq \xi_{\mathrm{U},\mathrm{G}}\right\}$,  $\mathcal{E}^{\text {Eve }} \triangleq\left\{\forall\left\|\Delta \mathbf{h}_{\text E}\right\|_{2} \leq \xi_{\mathrm{E},\mathrm{h}}, \right.$ $\left.\forall\left\|\Delta \mathbf{G}_{\text E}\right\|_{F} \leq \xi_{\mathrm{E},\mathrm{G}}\right\}$, constraints  \eqref{C3} and \eqref{C4} can be rewritten as
\begin{equation}
\label{linear_Alice}
\left|\left((\overline{\mathbf{h}}_{\mathrm{U}}\!+\!\Delta \mathbf{h}_{\mathrm{U}})\!+\!\mathbf{v}^{H} (\overline{\mathbf{G}}_{\mathrm{U}}\!+\!\Delta \mathbf{G}_{\mathrm{U}})\right) \mathbf{w}\right|^{2} \!\geq \! \beta_{\text{Alice}}\left(2^{\eta_{\text{U}}}\!-\!1\right), \mathcal{E}^{\text {Alice}},
\end{equation}
\begin{equation}
\label{linear_Eve}
\left|\left(\left(\overline{\mathbf{h}}_{\mathrm{E}}\!+\!\Delta \mathbf{h}_{\mathrm{E}}\right)\!+\!\mathbf{v}^{H}\left(\overline{\mathbf{G}}_{\mathrm{E}}\!+\!\Delta \mathbf{G}_{\mathrm{E}}\right)\right) \mathbf{w}\right|^{2} \!\leq \!\beta_{\text {Eve}}\left(2^{\eta_{\mathrm{E}}}\!-\!1\right), \mathcal{E}^{\text {Eve}},
\end{equation}
where $\beta_{\text{Alice}} = \sigma_{\mathrm{I}}^{2}\left\|\mathbf{v}^{H}\operatorname{diag}(\mathbf{h}_{\mathrm{IU}}^{H})\right\|^{2}+\sigma_{\mathrm{U}}^{2}$ and $\beta_{\text {Eve }}=\sigma_{\mathrm{E}}^{2}\left\|\mathbf{v}^{H}\operatorname{diag}(\mathbf{h}_{\mathrm{IE}}^{H})\right\|^{2}+\sigma_{\mathrm{E}}^{2}$ is the interference-plus-noises power at Alice and Eve, respectively.
In particular, the linear approximation of the useful signal power in \eqref{linear_Alice} is given in the following lemma.
\begin{lemma}
\label{lemma-1} Let $\mathbf{w}^{(k)}$ and $\mathbf{v}^{(k)}$ be the optimal solutions obtained at iteration $k$,  then the left side of the inequality \eqref{linear_Alice} $\left|\left[\left(\overline{\mathbf{h}}_{\mathrm{U}}+\Delta \mathbf{h}_{\mathrm{U}}\right)^{H}+\mathbf{v}^{H}\left(\overline{\mathbf{G}}_{\mathrm{U}}+\Delta \mathbf{G}_{\mathrm{U}}\right)\right] \mathbf{w}\right|^{2}$ is lower bounded linearly at $\left(\mathbf{w}^{(k)}, \mathbf{v}^{(k)}\right)$ as follows
\begin{align}
\label{lemma-1_eq}
\mathbf{x}_{1}^{H} \widetilde{\mathbf{A}} \mathbf{x}_{1}+2 \operatorname{Re}\left\{\widetilde{\mathbf{a}}_\mathrm{U}^{H} \mathbf{x}_{1}\right\}+\widetilde{a}_\mathrm{U},
\end{align}
\end{lemma}
where
%\begin{footnotesize}
\begin{equation}
\begin{aligned}
&\widetilde{\mathbf{A}}=\mathbf{C}+\mathbf{C}^{H}-\mathbf{Z},\\
&\widetilde{\mathbf{a}}_{\mathrm{U}}=\mathbf{c}_{\mathrm{1U}}+\mathbf{c}_{\mathrm{2U}}-\mathbf{z}_{\mathrm{U}},\\
&\widetilde{a}_{\mathrm{U}}=2 \operatorname{Re}\left\{c_{\mathrm{U}}\right\}-z_{\mathrm{U}},\\
&\mathbf{x}_{1}=\left[\Delta \mathbf{h}_{\mathrm{U}}^{H} \operatorname{vec}^{H}\left(\Delta \mathbf{G}_{\mathrm{U}}^{*}\right)\right]^{H}.
\end{aligned}
\end{equation}
%\end{footnotesize}
Note that $\mathbf{C}, \mathbf{Z}, \mathbf{c}_{\mathrm{1U}}, \mathbf{c}_{\mathrm{2U}}, \mathbf{z}_{\mathrm{U}}, c_{\mathrm{U}}$, and $z_{\mathrm{U}}$ are given by \eqref{Auxiliary_parameters} at the bottom of the page, respectively.
\begin{proof}Please refer to Appendix A.
\end{proof}

Thus, constraints \eqref{linear_Alice} is equivalently rewritten as
\begin{align}
\label{Alice_C}
&\mathbf{x}_{1}^{H} \widetilde{\mathbf{A}} \mathbf{x}_{1}+2 \operatorname{Re}\left\{\widetilde{\mathbf{a}}_{\mathrm{U}}^{H} \mathbf{x}_{1}\right\}+\widetilde{a}_{\mathrm{U}} \geq \beta_{\text{Alice}}\left(2^{\eta_{\text{U}}}-1\right), \mathcal{E}^{\text {Alice}}.
\end{align}
Similarly, constraint \eqref{linear_Eve} can be equivalently rewritten as
\begin{align}
\label{Eve_C}
&\mathbf{x}_{2}^{H} \widetilde{\mathbf{A}} \mathbf{x}_{2}+2 \operatorname{Re}\left\{\widetilde{\mathbf{a}}_{\mathrm{E}}^{H} \mathbf{x}_{2}\right\}+\widetilde{a}_{\mathrm{E}} \leq \beta_{\text{Eve}}\left(2^{\eta_{\mathrm{E}}}-1\right), \mathcal{E}^{\text{Eve}},
%\label{Eve_NOISE_C}
%&\sigma_{\mathrm{I}}^{2}\left\|\mathbf{h}_{\mathrm{IE}}^{H} \boldsymbol{\Theta}\right\|^{2}+\sigma_{\mathrm{E}}^{2} \geq \beta_{\text{Eve}}, \mathcal{E}^{\text{Eve}}
\end{align}
where
\begin{equation}
\begin{aligned}
&\widetilde{\mathbf{a}}_{\mathrm{E}}=\mathbf{c}_{\mathrm{1E}}+\mathbf{c}_{\mathrm{2E}}-\mathbf{z}_{\mathrm{E}},\\
&\widetilde{a}_{\mathrm{E}}=2 \operatorname{Re}\left\{c_{\mathrm{E}}\right\}-z_{\mathrm{E}},\\
&\mathbf{x}_{2}=\left[\Delta \mathbf{h}_{\mathrm{E}}^{H} \operatorname{vec}^{H}\left(\Delta \mathbf{G}_{\mathrm{E}}^{*}\right)\right]^{H},
\end{aligned}
\end{equation}
and $\mathbf{c}_{\mathrm{1E}}, \mathbf{c}_{\mathrm{2E}}, \mathbf{z}_{\mathrm{E}}, c_{\mathrm{E}}$, and $z_{\mathrm{E}}$ are also given by \eqref{Auxiliary_parameters}, respectively.
\begin{lemma}
\label{lemma-2} (S-Procedure \cite{S_lemma}): Let a function $f_{i}(\mathbf{x}), i \in$ $\{1,2\}, \mathbf{x} \in \mathbb{C}^{N \times 1}$, be defined as
\begin{align}
f_{i}(\mathbf{x})=\mathbf{x}^{H} \mathbf{B}_{i} \mathbf{x}+2 \operatorname{Re}\left\{\mathbf{b}_{i}^{\mathrm{H}} \mathbf{x}\right\}+b_{i}.
\end{align}
The condition $f_{1}(\mathbf{x}) \geq 0 \Rightarrow$ $f_{2}(\mathbf{x}) \geq 0$ holds if and only if there exist $\varpi \geq 0$ such that
\begin{equation}
\left[\begin{array}{ll}
\mathbf{B}_{2} & \mathbf{b}_{2} \\
\mathbf{b}_{2}^{H} & b_{2}
\end{array}\right]- \varpi\left[\begin{array}{ll}
\mathbf{B}_{1} & \mathbf{b}_{1} \\
\mathbf{b}_{1}^{H} & b_{1}
\end{array}\right] \succeq \mathbf{0}.
\end{equation}
Conversely, the condition $f_{1}(\mathbf{x}) \leq 0 \Rightarrow$ $f_{2}(\mathbf{x}) \leq 0$ holds if and only if there exist $\psi \geq 0$ such that
\begin{equation}
\psi\left[\begin{array}{ll}
\mathbf{B}_{1} & \mathbf{b}_{1} \\
\mathbf{b}_{1}^{H} & b_{1}
\end{array}\right]-\left[\begin{array}{ll}
\mathbf{B}_{2} & \mathbf{b}_{2} \\
\mathbf{b}_{2}^{H} & b_{2}
\end{array}\right] \succeq \mathbf{0}.
\end{equation}
\end{lemma}

In order to transform the constraints \eqref{linear_Alice} into LMIs by applying the Lemma 2, $\mathcal{E}^{\text{Alice}}$ is rewritten as the following quadratic expression
\begin{equation}
\mathcal{E}^{\text{Alice}} \triangleq\left\{\begin{array}{c}
\mathbf{x}_{1}^{H}\left[\begin{array}{cc}
\mathbf{I}_{N} & \mathbf{0} \\
\mathbf{0} & \mathbf{0}
\end{array}\right] \mathbf{x}_{1}-\xi_{\mathrm{U}, h}^{2} \leq 0, \\
\mathbf{x}_{1}^{H}\left[\begin{array}{cc}
\mathbf{0} & \mathbf{0} \\
\mathbf{0} & \mathbf{I}_{M N}
\end{array}\right] \mathbf{x}_{1}-\xi_{\mathrm{U}, G}^{2} \leq 0 .
\end{array}\right.
\end{equation}
Then, after introducing $\varpi_{\mathrm{U1}} \geq 0 $ and $\varpi_{\mathrm{U2}} \geq 0 $ as slack variables, constraints \eqref{Alice_C} can be transformed by Lemma 1 into the following equivalent LMIs as
\begin{equation}
\label{LMI_Alice}
\left[\begin{array}{cc}
\widetilde{\mathbf{A}}+\left[\begin{array}{cc}
\varpi_{\mathrm{U1}} \mathbf{I}_{N} & \mathbf{0} \\
\mathbf{0} & \varpi_{\mathrm{U2}} \mathbf{I}_{M N}
\end{array}\right] & \widetilde{\mathbf{a}}_{\mathrm{U}} \\
\widetilde{\mathbf{a}}_{\mathrm{U}}^{\mathrm{H}} & V_{\mathrm{U}}
\end{array}\right] \succeq \mathbf{0},
\end{equation}
where $V_{\mathrm{U}}\!=\!\widetilde{a}_\mathrm{U}\!-\!\beta_{\text{Alice}}\left(2^{R_{\mathrm{U}}}\!-\!1\right)\!-\!\varpi_{\mathrm{U1}} \xi_{\mathrm{U},\mathrm{h}}^{2}\!-\varpi_{\mathrm{U2}}\xi_{\mathrm{U},\mathrm{G}}^{2}$.
After introducing $\psi_{\mathrm{E1}} \geq 0 $ and $\psi_{\mathrm{E2}} \geq 0 $ as slack variables, constraints \eqref{Eve_C} can be transformed into the following equivalent LMIs as
\begin{equation}
\label{LMI_Eve}
\left[\begin{array}{cc}
\left[\begin{array}{cc}
\psi_{\mathrm{E}, \mathrm{h}} \mathbf{I}_{N} & \mathbf{0} \\
\mathbf{0} & \psi_{\mathrm{E2}} \mathbf{I}_{M N}
\end{array}\right] -\widetilde{\mathbf{A}}& -\widetilde{\mathbf{a}}_{\mathrm{E}} \\
-{\widetilde{\mathbf{a}}_{\mathrm{E}}^{\mathrm{H}}} & V_{\mathrm{E}}
\end{array}\right] \succeq \mathbf{0},
\end{equation}
where $V_{\mathrm{E}}\!=\!\beta_{\text {Eve}}\left(2^{R_{\mathrm{E}}}\!-1\right)\!-\widetilde{a}_{\mathrm{E}}\!-\psi_{\mathrm{E1}} \xi_{\mathrm{E}, \mathrm{h}}^{2}\!- \psi_{\mathrm{E2}} \xi_{\mathrm{E}, \mathrm{G}}^{2}.$

Subsequently, we transeform \eqref{C7} into
\begin{align}
\label{Active_S}
\left|\mathbf{v}^{H} \operatorname{diag}\left(\mathbf{H}_{\text{I}} \mathbf{w}\right)\right|^{2}+|\mathbf{v}|^{2} \sigma_{\text{I}}^{2} \leq P_{\text{F}}.
\end{align}
Then we adopt Schur's complement \cite{Convex} to equivalently convert \eqref{Active_S} into matrix inequalities as follows
\begin{equation}
\begin{aligned}
\label{Active_C}
&{\left[\begin{array}{cr}
P_{F}-|\mathbf{v}|^{2}\sigma_{1}^{2} %&\mathbf{\Theta} \mathbf{H}_{\mathrm{I}} \mathbf{w} \\
&\mathbf{v}^{H} \operatorname{diag}\left(\mathbf{H}_{\text{I}} \mathbf{w}\right)\\
\operatorname{diag}\left(\mathbf{H}_{\text{I}} \mathbf{w}\right)^{H} \mathbf{v}  &\mathbf{I}
%\mathbf{w}^{H} \mathbf{H}_{\mathrm{I}}^{H} \mathbf{\Theta}^{H}  & \mathbf{I}
\end{array}\right] \succeq \mathbf{0}}.
\end{aligned}
\end{equation}

 Based on the above discussions, for given $\mathbf{v}$, the subproblem of $\mathbf{w}$ is given by
\begin{align}
\label{P3}
\notag \min _{\mathbf{w}, \boldsymbol{\varpi}_{\mathrm{U1}}, \atop \boldsymbol{\varpi}_{\mathrm{U2}},  \boldsymbol{\psi}_{\mathrm{E1}}, \boldsymbol{\psi}_{\mathrm{E2}}} \ &\|\mathbf{w}\|^{2} \\
\text { s.t. }  &\eqref{C1},\eqref{LMI_Alice},\eqref{LMI_Eve},\eqref{Active_C}.
\end{align}
Problem \eqref{P3} is a convex problem and can be directly solved by the CVX tool.

Then, for given $\mathbf{w}$, the subproblem of $\mathbf{v}$ is a feasibility-check problem. According to \cite{f_check} and in order to improve the converged solution in the optimization of $\mathbf{v}$, the useful signal power inequalities in \eqref{linear_Alice} are modified by introducing slack variables $\alpha_{\mathrm{U}} \geq 0 $ and recast as
\begin{align}
\label{fc_Alice}
\notag &\left|\left((\overline{\mathbf{h}}_{\mathrm{U}}+\Delta \mathbf{h}_{\mathrm{U}})+\mathbf{v}^{H} (\overline{\mathbf{G}}_{\mathrm{U}}+\Delta \mathbf{G}_{\mathrm{U}})\right) \mathbf{w}\right|^{2} \\
&\geq \beta_{\text{Alice}}\left(2^{\eta_{\text{U}}}-1\right)+\alpha_{\text{U}}, \mathcal{E}^{\text {Alice}}.
\end{align}
Subsequently, the LMIs \eqref{LMI_Alice} are modified as
\begin{equation}
\label{fc_LMI_Alice}
\left[\begin{array}{cc}
\widetilde{\mathbf{A}}+\left[\begin{array}{cc}
\varpi_{\mathrm{U}, \mathrm{h}} \mathbf{I}_{N} & \mathbf{0} \\
\mathbf{0} & \varpi_{\mathrm{U}, \mathrm{G}} \mathbf{I}_{M N}
\end{array}\right] & \widetilde{\mathbf{a}}_{\mathrm{U}} \\
\widetilde{\mathbf{a}}_{\mathrm{U}}^{\mathrm{H}} & V_{\mathrm{U}}-\alpha_{\text{U}}
\end{array}\right] \succeq \mathbf{0},
\end{equation}
the Eve signal power inequalities in \eqref{linear_Eve} are modified by introducing slack variables $\alpha_{\mathrm{E}} \geq 0 $ and recast as
\begin{align}
\label{fc_Eve}
\notag&\left|\left(\left(\overline{\mathbf{h}}_{\mathrm{E}}+\Delta \mathbf{h}_{\mathrm{E}}\right)+\mathbf{v}^{H}\left(\overline{\mathbf{G}}_{\mathrm{E}}+\Delta \mathbf{G}_{\mathrm{E}}\right)\right) \mathbf{w}\right|^{2} \\
&\leq \beta_{\text{Eve}}\left(2^{\eta_{\mathrm{E}}}-1\right)-\alpha_{\text{E}}, \mathcal{E}^{\text{Eve}},
%\label{Eve_NOISE_C}
%&\sigma_{\mathrm{I}}^{2}\left\|\mathbf{h}_{\mathrm{IE}}^{H} \boldsymbol{\Theta}\right\|^{2}+\sigma_{\mathrm{E}}^{2} \geq \beta_{\text{Eve}}, \mathcal{E}^{\text{Eve}}
\end{align}
and the LMIs \eqref{LMI_Alice} are modified as
\begin{equation}
\label{fc_LMI_Eve}
\left[\begin{array}{cc}
\left[\begin{array}{cc}
\varpi_{\mathrm{E}, \mathrm{h}} \mathbf{I}_{N} & \mathbf{0} \\
\mathbf{0} & \varpi_{\mathrm{E}, \mathrm{G}} \mathbf{I}_{M N}
\end{array}\right] -\widetilde{\mathbf{A}}& -\widetilde{\mathbf{a}}_{\mathrm{E}} \\
-{\widetilde{\mathbf{a}}_{\mathrm{E}}^{H}} & V_{\mathrm{E}}-\alpha_{\text{E}}
\end{array}\right] \succeq \mathbf{0}.
\end{equation}
Let $\mathbf{h}_{\text{g}}=\mathbf{H}_{\mathrm{I}}\mathbf{w}$, and $\mathbf{F}=\operatorname{diag}\left(\left[\left|h_{\text{g1}}\right|^{2}, \ldots,\left|h_{\text{gM}}\right|^{2}\right]\right)+\sigma_{2}^{2} \mathbf{I}_{\text{M}}$, \eqref{F} can be expressed as $\mathbf{v}^{\mathrm{H}} \mathbf{F v} \leq P_{\text {F}}$.
Thus, the sub-problem of $\mathbf{v}$ can be formulated as
\begin{equation}
\label{P4}
\begin{aligned}
\max _{\mathbf{v}, \boldsymbol{\varpi}_{\mathrm{U1}}, \boldsymbol{\varpi}_{\mathrm{U2}}, \atop  \boldsymbol{\psi}_{\mathrm{E1}}, \boldsymbol{\psi}_{\mathrm{E2}}, \boldsymbol{\alpha}_{\mathrm{U}},\boldsymbol{\alpha}_{\mathrm{E}}} \ &\alpha_{\text{U}} + \alpha_{\text{E}} \\   %  \ 表示一个空格
\text { s.t. } \ &\eqref{fc_LMI_Alice},\eqref{fc_LMI_Eve},\\
%&\left|\mathbf{v}^{H} \operatorname{diag}\left(\mathbf{H}_{\text{I}} \mathbf{w}\right)\right|^{2}+|\mathbf{v}|^{2} \sigma_{\text{I}}^{2} \leq P_{\text{F}}.
&\mathbf{v}^{\mathrm{H}} \mathbf{F v} \leq P_{\text {F}}, \\
& |\mathbf{v}[m]| \leq \tau_{\text{max}}
\end{aligned}
\end{equation}
Problem \eqref{P4} is a convex optimization problem which also can be solved efficiently and optimally with the tool CVX.
%\newcounter{mytempeqncnt2}  %图片占双栏置顶
%\begin{figure*}[!t]
%\normalsize

\begin{equation}
\label{Auxiliary_parameters}
\begin{aligned}
\notag
&\mathbf{C}=\left[\begin{array}{c}
\mathbf{w}^{(k)} \\
\mathbf{w}^{(k)} \otimes \mathbf{v}^{(k), *}
\end{array}\right]  \left[\mathbf{w}^{H} \mathbf{w}^{H} \otimes \mathbf{v}^{T}\right],\\
&\mathbf{Z}=\left[\begin{array}{c}
\mathbf{w}^{(k)} \\
\mathbf{w}^{(k)} \otimes \mathbf{v}^{(k), *}
\end{array}\right]\!\left[\mathbf{w}^{(k), H} \mathbf{w}^{(k), H} \otimes \mathbf{v}^{(k), T}\right],\\
&\mathbf{c}_{\mathrm{1U}}=\left[\begin{array}{c}
\mathbf{w} \mathbf{w}^{(k), H}\left(\overline{\mathbf{h}}_{\mathrm{U}}+\overline{\mathbf{G}}_{\mathrm{U}}^{H} \mathbf{v}^{(k)}\right) \\
\operatorname{vec}^{*}\left(\mathbf{v}\left(\overline{\mathbf{h}}_{\mathrm{U}}^{H}+\mathbf{v}^{(k), H} \overline{\mathbf{G}}_{\mathrm{U}}\right) \mathbf{w}^{(k)} \mathbf{w}^{H}\right)
\end{array}\right],\\
&\mathbf{c}_{\mathrm{2U}}=\left[\begin{array}{c}
\mathbf{w}^{(k)} \mathbf{w}^{H}\left(\overline{\mathbf{h}}_{\mathrm{U}}+\overline{\mathbf{G}}_{\mathrm{U}}^{H} \mathbf{v}\right) \\
\operatorname{vec}^{*}\left(\mathbf{v}^{(k)}\left(\overline{\mathbf{h}}_{\mathrm{U}}^{H}+\mathbf{v}^{H} \overline{\mathbf{G}}_{\mathrm{U}}\right) \mathbf{w} \mathbf{w}^{(k), H}\right)
\end{array}\right],\\
&\mathbf{z}_{\mathrm{U}}=\left[\begin{array}{c}
\mathbf{w}^{(k)} \mathbf{w}^{(k), H}\left(\overline{\mathbf{h}}_{\mathrm{U}}+\overline{\mathbf{G}}_{\mathrm{U}}^{H} \mathbf{v}^{(k)}\right) \\
\operatorname{vec}^{*}\left(\mathbf{v}^{(k)}\left(\overline{\mathbf{h}}_{\mathrm{U}}^{H}+\mathbf{v}^{(k), H} \overline{\mathbf{G}}_{\mathrm{U}}\right) \mathbf{w}^{(k)} \mathbf{w}^{(k), H}\right)
\end{array}\right],\\
&c_{\mathrm{U}}=\left(\overline{\mathbf{h}}_{\mathrm{U}}^{H}+\mathbf{v}^{(k), H} \overline{\mathbf{G}}_{\mathrm{U}}\right) \mathbf{w}^{(k)} \mathbf{w}^{H}\left(\overline{\mathbf{h}}_{\mathrm{U}}+\overline{\mathbf{G}}_{\mathrm{U}}^{H} \mathbf{v}\right),\\
&z_{\mathrm{U}}=\left(\overline{\mathbf{h}}_{\mathrm{U}}^{H}+\mathbf{v}^{(k), H} \overline{\mathbf{G}}_{\mathrm{U}}\right) \mathbf{w}^{(k)} \mathbf{w}^{(k), H}\left(\overline{\mathbf{h}}_{\mathrm{U}}+\overline{\mathbf{G}}_{\mathrm{U}}^{H} \mathbf{v}^{(k)}\right),\\
&\mathbf{c}_{\mathrm{1E}}=\left[\begin{array}{c}
\mathbf{w} \mathbf{w}^{(k), H}\left(\overline{\mathbf{h}}_{\mathrm{E}}+\overline{\mathbf{G}}_{\mathrm{E}}^{H} \mathbf{v}^{(k)}\right) \\
\operatorname{vec}^{*}\left(\mathbf{v}\left(\overline{\mathbf{h}}_{\mathrm{E}}^{H}+\mathbf{v}^{(k), H} \overline{\mathbf{G}}_{\mathrm{E}}\right) \mathbf{w}^{(k)} \mathbf{w}^{H}\right)
\end{array}\right],\\
&\mathbf{c}_{\mathrm{2E}}=\left[\begin{array}{c}
\mathbf{w}^{(k)} \mathbf{w}^{H}\left(\overline{\mathbf{h}}_{\mathrm{E}}+\overline{\mathbf{G}}_{\mathrm{E}}^{H} \mathbf{v}\right) \\
\operatorname{vec}^{*}\left(\mathbf{v}^{(k)}\left(\overline{\mathbf{h}}_{\mathrm{E}}^{H}+\mathbf{v}^{H} \overline{\mathbf{G}}_{\mathrm{E}}\right) \mathbf{w} \mathbf{w}^{(k), H}\right)
\end{array}\right],\\
\end{aligned}
\end{equation}
\begin{equation}
\begin{aligned}
&\mathbf{z}_{\mathrm{E}}=\left[\begin{array}{c}
\mathbf{w}^{(k)} \mathbf{w}^{(k), H}\left(\overline{\mathbf{h}}_{\mathrm{E}}+\overline{\mathbf{G}}_{\mathrm{E}}^{H} \mathbf{v}^{(k)}\right) \\
\operatorname{vec}^{*}\left(\mathbf{v}^{(k)}\left(\overline{\mathbf{h}}_{\mathrm{E}}^{H}+\mathbf{v}^{(k), H} \overline{\mathbf{G}}_{\mathrm{E}}\right) \mathbf{w}^{(k)} \mathbf{w}^{(k), H}\right)
\end{array}\right],\\
&c_{\mathrm{E}}=\left(\overline{\mathbf{h}}_{\mathrm{E}}^{H}+\mathbf{v}^{(k), H} \overline{\mathbf{G}}_{\mathrm{E}}\right) \mathbf{w}^{(k)} \mathbf{w}^{H}\left(\overline{\mathbf{h}}_{\mathrm{E}}+\overline{\mathbf{G}}_{\mathrm{E}}^{H} \mathbf{v}\right),\\
&z_{\mathrm{E}}=\left(\overline{\mathbf{h}}_{\mathrm{E}}^{H}+\mathbf{v}^{(k), H} \overline{\mathbf{G}}_{\mathrm{E}}\right) \mathbf{w}^{(k)} \mathbf{w}^{(k), H}\left(\overline{\mathbf{h}}_{\mathrm{E}}+\overline{\mathbf{G}}_{\mathrm{E}}^{H} \mathbf{v}^{(k)}\right).\\
\end{aligned}
\end{equation}

%\hrulefill
%\vspace*{4pt}
%\end{figure*}

We summarize the proposed AO algorithm in Algorithm 1.
\begin{algorithm}[htbp]
	\caption{The proposed AO algorithm for secure and robust communication with jittering UAV}
	\label{2}
{\bf Initialize:} % 算法的输入， \hspace*{0.02in} 用来控制位置，同时利用 \\ 进行换行
    Set $k=0$, and initialize $\mathbf{v}^{(0)}, \mathbf{w}^{(0)}$. Compute the objective function value of problem \eqref{P1} (i.e.,
the transmit power) as $p(\mathbf{v}^{(0)}, \mathbf{w}^{(0)})$. Give the error tolerance $\epsilon$.\\
{\bf Repeat:} % 算法的结果输出
\begin{algorithmic}
  \State 1.  Solve the problem \eqref{P3} to obtain $\mathbf{w}^{(k+1)}$ for given $\mathbf{v}^{(k)}$.
  \State 2.  Solve the problems \eqref{P4} to obtain $\mathbf{v}^{(k+1)}$ for given $\mathbf{w}^{(k+1)}$.
  \State 3.  With given $\mathbf{v}^{(k+1)}$ and $\mathbf{w}^{(k+1)}$, compute the objective function value of problem \eqref{P1} as $p\left(\mathbf{v}^{(k+1)},\mathbf{w}^{(k+1)}\right)$.
  \State 4.  $k:=k+1$.
  \end{algorithmic}
{\bf Until:} $\left|p\left(\mathbf{v}^{(k)},\mathbf{w}^{(k)}\right)\!-\!p\left(\mathbf{v}^{(k\!-\!1)},\mathbf{w}^{(k\!-\!1)} \right)\right| / p\left(\mathbf{v}^{(k\!-\!1)}, \mathbf{w}^{(k\!-\!1)}\right)$\\ $<\epsilon$ or problem \eqref{P4} becomes infeasible.
\end{algorithm}

\section{Simulation Results}
\begin{figure}[tp]
\centering
\subfigure[] %子图片标题
{\includegraphics[width=3.3in]{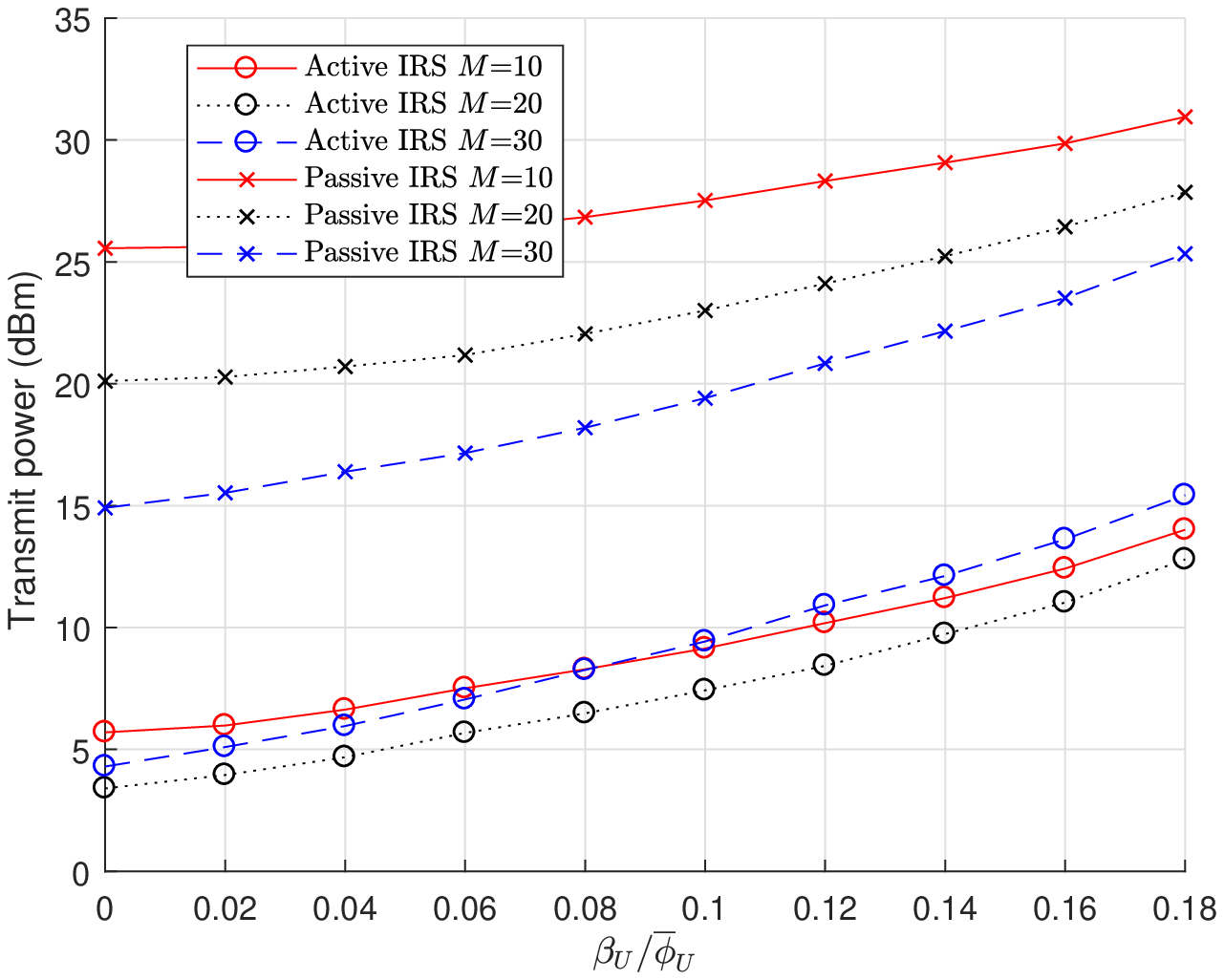}} %[图片大小]{图片路径}
\subfigure[]
{\includegraphics[width=3.3in]{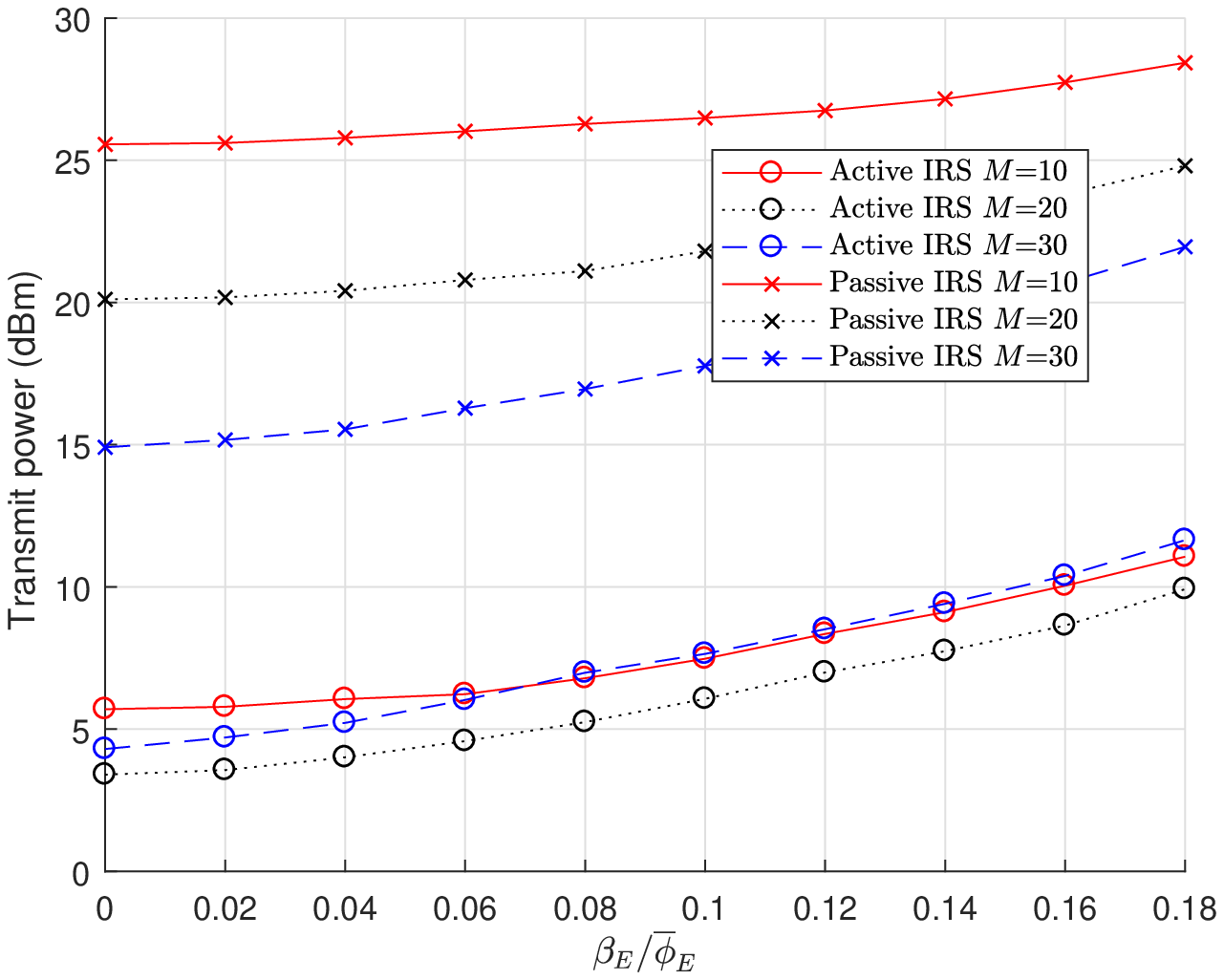}}
\caption{Comparison between different $M$ versus the variations of AOD.} %图片标题
\label{UAV_M}  %图片交叉引用时的标签
\end{figure}

\begin{figure}[tp]
\centering
\subfigure[] %子图片标题
{\includegraphics[width=3.3in]{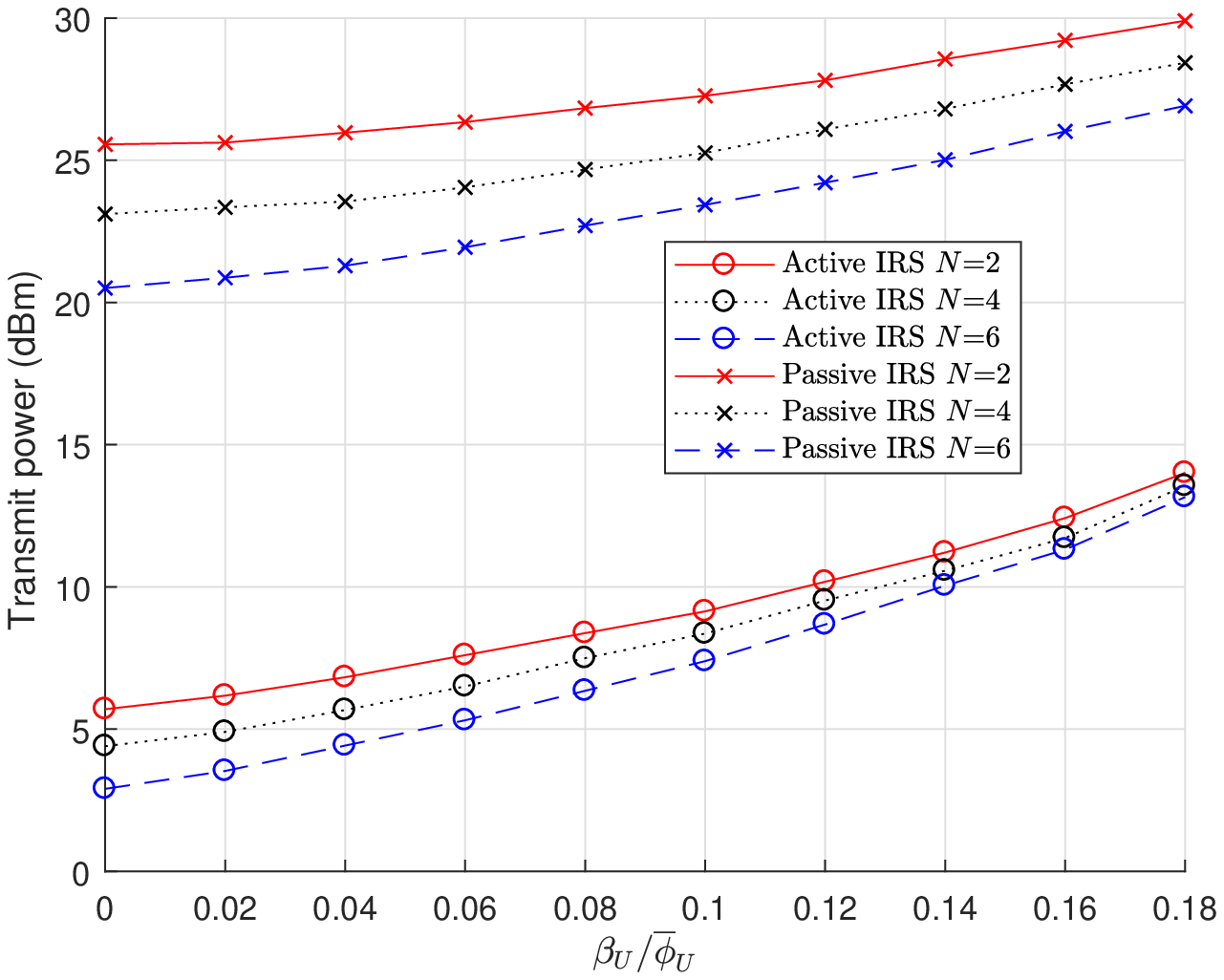}} %[图片大小]{图片路径}
\subfigure[]
{\includegraphics[width=3.3in]{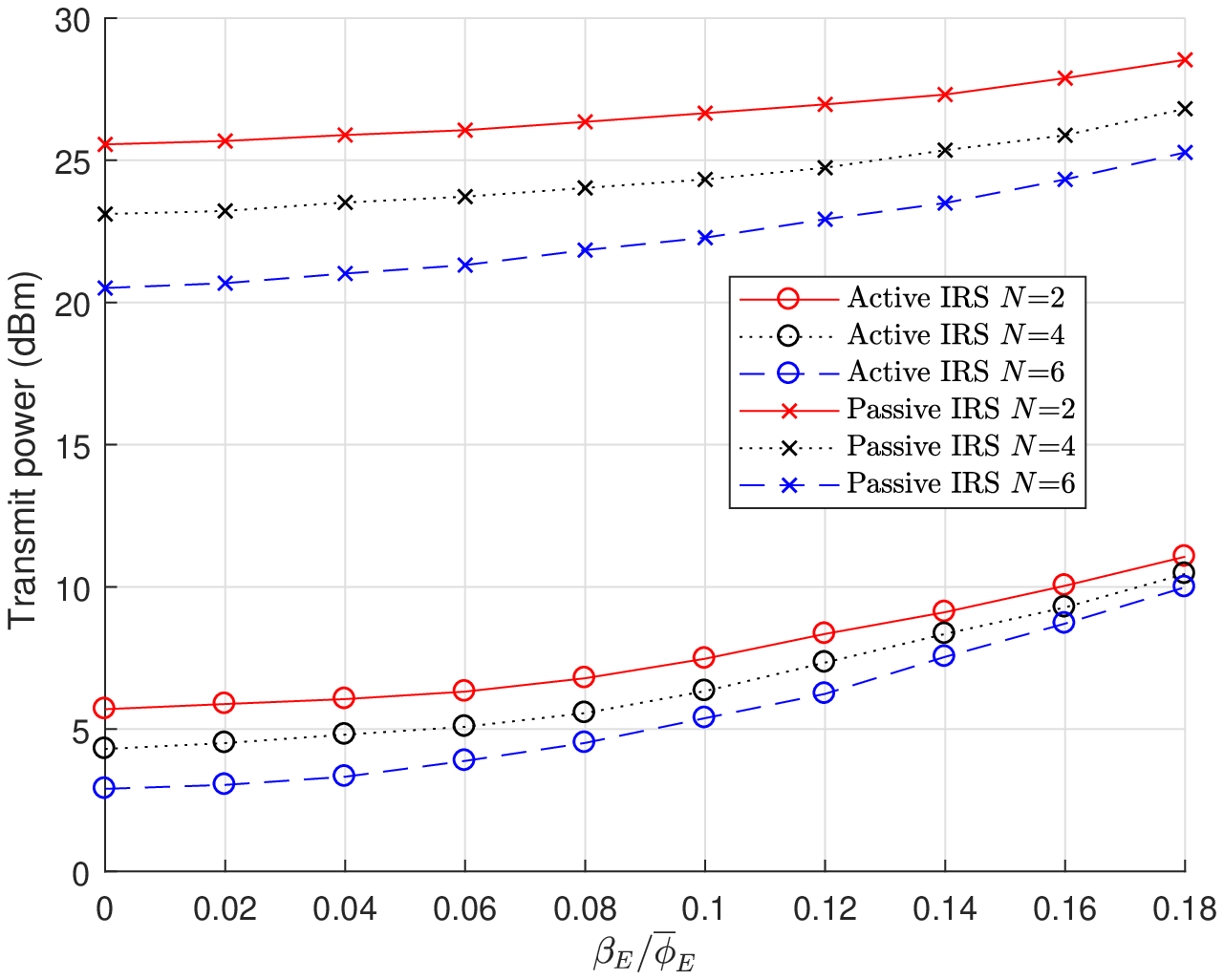}}
\caption{Comparison between different $N$ versus the variations of AOD.} %图片标题
\label{UAV_N}  %图片交叉引用时的标签
\end{figure}

\begin{figure}[htbp]
\centerline{\includegraphics[width=3.3in]{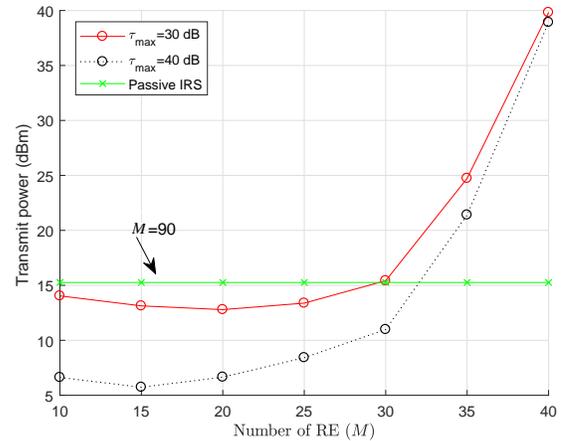}}
	\caption{Comparison between transmission power versus the active IRS's maximum amplification factor $\tau_{\text{max}}$.}
\label{M_V}
\end{figure}

\begin{figure}[htbp]
\centerline{\includegraphics[width=3.1in]{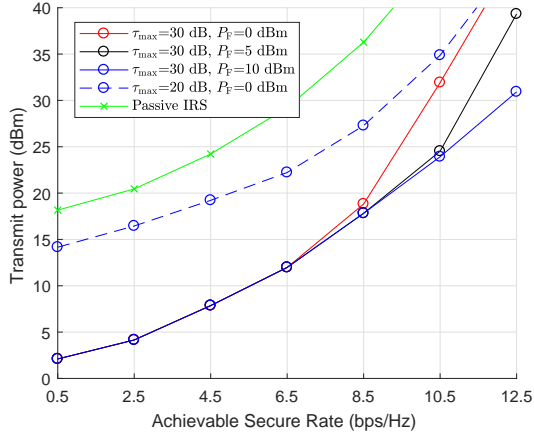}}
	\caption{Comparison between different ASR versus the active IRS's maximum amplification power $P_{\text{F}}$.}
\label{Rate}
\end{figure}

\begin{figure}[htbp]
\centerline{\includegraphics[width=3.3in]{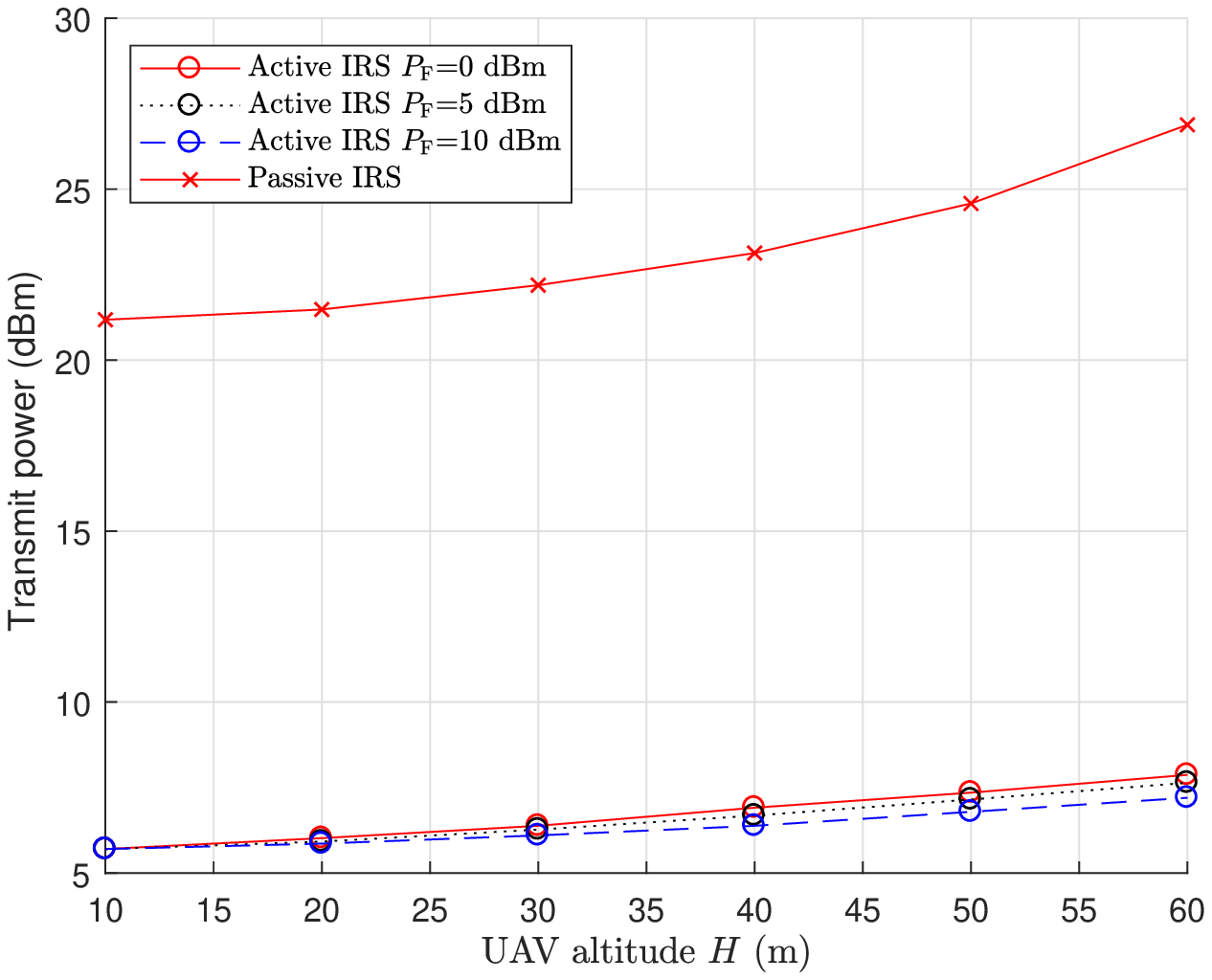}}
	\caption{Comparison between different UAV altitude versus the active IRS's maximum amplification power $P_{\text{F}}$.}
\label{UAV_H}
\end{figure}

In this section, the simulation results are presented to demonstrate the secrecy performance of the proposed scheme. Note that transmission power in Fig.\ref{UAV_M}, Fig.\ref{UAV_N}, Fig.\ref{M_V}, and Fig.\ref{UAV_H} is optimized under the given ASR constraints to ensure secure UAV communication requirements. Fig.\ref{Rate} analyzes the improvement of the power efficiency of the proposed scheme at different ASR requirements. In addition, we focus on analyzing the transmission power optimization with reference to several important parameters, i.e., channel uncertainty caused by UAV jittering, the number of RE, the power amplification characteristics of active IRS, and UAV's height. Without specification, the coordinations of UAV, IRS, Alice and Eve are (10, 20, 10), (10, 0, 10), (20, 20, 0) and (10, 40, 0), respectively. It is assumed that UBS are equipped with $N = 2$ transmit antennas, $P_{\text{F}}= 10$ dBm, $\tau_{\text{max}}=30$ dB, $\eta_{\mathrm{U}}=4.5$, $\eta_{\mathrm{E}}=1$, unless otherwise specified. The Rician $K$-factors $K_{\mathrm{G}}$ and $K_{\mathrm{E}}$ are given by $K_{*}=a e^{b\left(\frac{\pi}{2}-\phi_{*}\right)}$ with $*=\{\mathrm{U}, \mathrm{E}, \mathrm{I}\}$ with environment parameters $a=5, b=\frac{2}{\pi} \ln 3$ as in \cite{Rician}. Meanwhile, the maximum transmission power $P_{\text {peak}}=40$ dBm , the path loss factors $A_{\mathrm{L}}=-2.14$ dB, $A_{\mathrm{N}}=-3.14$ dB, the path loss exponents $\alpha_{\mathrm{L}}=2.09, \alpha_{\mathrm{N}}=3.75 $ \cite{EE_UAV}, respectively. The noise power are set to be $\sigma_{I}^{2} =\sigma_{U}^{2} = \sigma_{E}^{2} =-10$ dBm, unless otherwise specified. ${\beta_{\text{E}1}}= {\beta_{\text{E}2}}=\frac{1}{2}{\beta_{\text{E}}}$ and ${\beta_{\text{U}1}}= {\beta_{\text{U}2}}=\frac{1}{2}{\beta_{\text{U}}}$ measure the relative amount of CSI uncertainties. In addition, we set error tolerance $\epsilon=10^{-4}$ in Algorithm 1. To demonstrate the advantage of the proposed scheme (denoted as ``Active IRS"), we also compare the results with the benchmark schemes, e.i., similar scheme assisted by passive IRS (denoted as ``Passive IRS").

\subsection{Impact of Channel Uncertainty}
%${\beta_{\mathrm{U}}}$ and ${\beta_{\mathrm{E}}}$}
To evaluate the impact of channel uncertainty caused by UAV jittering for active IRS-assisted UAV secure communication, Fig. \ref{UAV_M}(a) and Fig. \ref{UAV_M}(b) show the transmission power w.r.t the variations of Alice's AOD and Eve's AOD (e.i., the ratio between the maximum variation and the elevation AOD) under different number of elements $M$, respectively. Because the analysis of elevation angle $\phi$ is similar to that of azimuth angle $\omega$, here we take elevation angle as an example. Note that, $\frac{\beta_{\mathrm{E}}}{\bar{\phi}_{\mathrm{E}}}=0.04$ for Fig. \ref{UAV_M}(a), and $\frac{\beta_{\mathrm{U}}}{\bar{\phi}_{\mathrm{U}}}=0.02$ for Fig. \ref{UAV_M}(b). It can be seen that the transmission power at UBS increases with the increase of AOD uncertainty, which means that the more drastic of UAV jittering, the more transmission power is required to satisfy the effectiveness of the legitimate link and the security requirement of the confidential signal. In addition, the active IRS solution can achieve a significant improvement of secrecy performance with only a small number of elements, that is, reduce the UAV's transmission power under the same ASR. However, the equivalent noise due to the CSI deviation will be amplified by active IRS, thus resulting in that the secrecy performance gain from increasing the number of RE is not as significant as with passive IRS. Furthermore, unlike the passive IRS solution, the secrecy performance of active IRS solution will not continue to be improved as $M$ increases. If the number of elements is too large, the secrecy performance will decrease instead. Because active IRS is limited by the given amplification gain $P_{\text{F}}$, the amplification factor will decrease as the number of elements continue to increase. In conclusion, compared to passive IRS, robust beamforming design is more important for active IRS-assisted UAV secure communication.

Under the same settings as Figure. \ref{UAV_M} and assuming $M$ = 10, Fig. \ref{UAV_N}(a) and Fig. \ref{UAV_N}(b) show the transmission power w.r.t the variations of Alice's AOD and Eve's AOD under different number of transmitting antennas $N$, respectively. Different from the variation of $M$, the transmitting power of both the active IRS scheme and the passive IRS scheme continues to decrease as $N$ increases as the amplified noise at the active IRS is independent of the $N$. Besides, the secrecy performance gain brought by the increase of $N$ is more significant than that of $M$, e.i., the transmission power can be greatly reduced with a small increase in the number of transmitting antennas. This is due to the fact that the antenna beams become more concentrated in the targeted direction with more antennas equipped at the UBS, and thus less total transmission power is required. However, for active IRS, when the channel uncertainty increases, the secrecy performance gain will also gradually decrease as $N$ increases due to the more severe noise caused by UAV jittering will be amplified.
\subsection{Impact of the Number of RE}
Assuming that $\frac{\beta_{\mathrm{U}}}{\bar{\phi}_{\mathrm{U}}} = 0.18$ and $\frac{\beta_{\mathrm{E}}}{\bar{\phi}_{\mathrm{E}}}=0.04$, Fig. \ref{M_V} compares the transmission power of the IRS-assisted system in terms of the number of RE. In this setup, the passive IRS can employ 90 REs at most to reflect the incident signal. However, even with 90 REs, the transmission power with the passive IRS is more than that with the active IRS equipped with $M^{*}$ (optimum $M$) antennas, as the active IRS can directly amplify the incident signal. On the contrary, the active IRS only takes few REs (15-20 or so) to reach its optimal secrecy performance under different amplitude gains. With the less REs, the active IRS is of smaller surface size and more suitable to the load-limited UAV secure communication. Similar to Fig. \ref{UAV_M}, for given maximum amplification gain $\tau_{\text{max}}$, it is observed that with the increase in $M$, the transmission power for the active IRS first decreases and then increases. When $M$ is larger (over 30 or 35), the transmission power of the active IRS solution is higher than that of the passive IRS solution. Because the active IRS-assisted system can benefit more from the increase in $M$ at first, but when $M$ over its optimum value, the transmission power will increase, as more REs reduce the power left for amplification. Besides, the active IRS can reduce transmission power by increasing not only the number of REs but also the maximum amplification gain $\tau_{\text{max}}$. For larger $M$, the gain brought by higher $\tau_{\text{max}}$ will be more limited, as $\tau_{\text{max}}$ is also constrained with less amplification power. It alerts us that the tradeoff between the number of RE $M$ and the maximum amplification gain $\tau_{\text{max}}$ is very important. Undoubtedly, under the premise of meeting the same ASR, the proposed scheme with smaller surface size and lower power consumption has wide application value.
\subsection{Impact of the Maximum Amplification Gain and the Maximum Amplification Power}
%$\tau_{\mathrm{max}}$ and the Naximum Amplification Power $P_{\mathrm{F}}$}
%To evaluate the impact of constraints of effectiveness of the legitimate link and security of the confidential message.
To provide a basis for the design of active IRS for power optimization problems under different ASR requirements. Fig. \ref{Rate} show the transmission power w.r.t the different ASR thresholds (Note that, $\eta_{\mathrm{E}}=1$) under different maximum amplification power $P_{\text{F}}$. Assuming that $\frac{\beta_{\mathrm{E}}}{\bar{\phi}_{\mathrm{E}}}=0.04$, and $\frac{\beta_{\mathrm{U}}}{\bar{\phi}_{\mathrm{U}}}=0.06, M=20$. For a given $\tau_{\text{max}}=30$ dB, it is observed that when the required ASR is small, the IRS-assisted system with different amplification power budget has almost the same transmit power, which implies that the amplification power budget constraint is inactive for the weak transmit power. Therefore, for case with low security rate requirements ($R_{\text{sec}}< 6.5$ bps/Hz), only a small amplification power $P_{\text{F}}$ needs to be reserved for the IRS. However, as the required ASR increases, the curves with different $P_{\text{F}}$ are distinguished from the other, and the increase of the transmission power becomes dramatic. Because the constraint will be active as the transmission power increases, resulting in less amplification gain at the active IRS. It means that sufficient amplification power needs to be provided to meet the requirement of high ASR. Meanwhile, the active IRS with larger $\tau_{\text{max}}$ can provide more amplification gain with the same $P_{\text{F}}$, thereby reducing the transmission power for the lower ASR requirements. But when ASR gradually increase, the transmission power with different $\tau_{\text{max}}$ tend to be consistent, because the gradually increasing transmission power will reduce the actual amplification gain of the active IRS, thus avoiding the difference of $\tau_{\text{max}}$. Therefore, when designing the active IRS-assisted system, the IRS has to reconfigure its maximum amplification gain according to the required ASR.
\subsection{Impact of UAV's Height}
%: H}
 Fig.\ref{UAV_H} show the transmission power w.r.t the different UAV altitude under different maximum amplification power $P_{\text{F}}$. Assuming that $\frac{\beta_{\mathrm{E}}}{\bar{\phi}_{\mathrm{E}}}=0.04$, and $\frac{\beta_{\mathrm{U}}}{\bar{\phi}_{\mathrm{U}}}=0.06, M=20$. The passive IRS solution requires more transmit power when the UAV's flight altitude increases due to path loss while the active IRS scheme has almost the same performance under different rated amplification power constraints, that is, only a small amplification power is required to greatly reduce the transmission power required by the UBS. Because the larger the distance between the UBS and the IRS, the weaker the power of the incident signal, and then, according to \eqref{F}, the active IRS can provide more amplification gain to compensate for the attenuation caused by the double-chain path loss. Therefore, even for UAVs with high flying altitude, the transmission power at UBS can also be greatly reduced, which shows active IRS solutions can also provide better secrecy performance for long-distance UAV transmission.
\section{Conclusion}
In this paper, we proposed a novel active IRS assisted robust secure beamforming designing for UAV communication considering the impact of UAV jittering, aming to reduce the transmission power at UBS while ensuring ASR. Specifically, with appropriately designed reflecting coefficient, the active IRS can create enhanced reflective paths for legitimate users by amplifying the incident signal, while simultaneously reducing the eavesdropping of user's information by illegal eavesdroppers. Firstly, linear approximation applying by Taylor expansion is used to characterize the impact of the variations of elevation AOD and azimuth AOD on the channel response. Based on it, a power minimization problem is formulated with constraints for the effectiveness of legitimate link and the security requirement of confidential message, where the precoding matrix and the active IRS's reflecting coefficient were jointly optimized. Secondly, S-procedure, Schur's complement procedure, and slack variables were adopted to transform the original highly non-convex problem into a linear convex problem, and the AO algorithm was introduced to alternately optimize the reflecting coefficient matrix and transmit beamformer matrix. Simulation results showed that the proposed robust scheme assisted by active IRS achieves superior performance compared with the passive IRS scheme, even in the complex situation of UAV's power limitation, severe jittering, and high flight altitude. In general, it reveals the potential of active IRS in secure UAV communication and the impact of the channel uncertainty caused by UAV jittering.

\newcounter{mytempeqncnt2}  %图片占双栏置顶
\begin{figure*}[!t]
\normalsize
\begin{equation}
\begin{aligned}
\label{Taylor_3}
&\left[\left(\overline{\mathbf{h}}^{H}+\Delta \mathbf{h}^{H}\right)+\mathbf{v}^{(k), H}\left(\overline{\mathbf{G}}+\Delta \mathbf{G}\right)\right] \mathbf{w}^{(k)} \mathbf{w}^{H}\left[\left(\overline{\mathbf{h}}+\Delta \mathbf{h}\right)+\left(\overline{\mathbf{G}}^{H}+\Delta \mathbf{G}^{H}\right) \mathbf{v}\right] \\
=&\left(\overline{\mathbf{h}}^{H}+\mathbf{v}^{(k), H} \overline{\mathbf{G}}\right) \mathbf{w}^{(k)} \mathbf{w}^{H}\left(\overline{\mathbf{h}}+\overline{\mathbf{G}}^{H} \mathbf{v}\right)+\left(\overline{\mathbf{h}}^{H}+\mathbf{v}^{(n), H} \overline{\mathbf{G}}\right) \mathbf{w}^{(k)} \mathbf{w}^{H}\left(\Delta \mathbf{h}+\Delta \mathbf{G}^{H} \mathbf{v}\right) \\
&+\left(\Delta \mathbf{h}^{H}+\mathbf{v}^{(k), H} \Delta \mathbf{G}\right) \mathbf{w}^{(k)} \mathbf{w}^{H}\left(\overline{\mathbf{h}}+\overline{\mathbf{G}}^{H} \mathbf{v}\right)+\left(\Delta \mathbf{h}^{H}+\mathbf{v}^{(k), H} \Delta \mathbf{G}\right) \mathbf{w}^{(k)} \mathbf{w}^{H}\left(\Delta \mathbf{h}+\Delta \mathbf{G}^{H} \mathbf{v}\right) \\
=&\left(\overline{\mathbf{h}}^{H}+\mathbf{v}^{(k), H} \overline{\mathbf{G}}\right) \mathbf{w}^{(k)} \mathbf{w}^{H}\left(\overline{\mathbf{h}}+\widehat{\mathbf{G}}^{H} \mathbf{v}\right)+\left(\overline{\mathbf{h}}^{H}+\mathbf{v}^{(k), H} \overline{\mathbf{G}}\right) \mathbf{w}^{(k)} \mathbf{w}^{H} \Delta \mathbf{h} \\
&+\operatorname{vec}^{H}\left(\Delta \mathbf{G}\right) \operatorname{vec}\left(\mathbf{v}\left(\overline{\mathbf{h}}^{H}+\mathbf{v}^{(k), H} \overline{\mathbf{G}}\right) \mathbf{w}^{(k)} \mathbf{w}^{H}\right)+\Delta \mathbf{h}^{H} \mathbf{w}^{(k)} \mathbf{w}^{H}\left(\overline{\mathbf{h}}+\overline{\mathbf{G}}^{H} \mathbf{v}\right)+\Delta \mathbf{h}^{H} \mathbf{w}^{(k)} \mathbf{w}^{H} \Delta \mathbf{h} \\
&+\operatorname{vec}^{H}\left(\mathbf{v}^{(k)}\left(\overline{\mathbf{h}}+\mathbf{v}^{H} \overline{\mathbf{G}}^{H}\right) \mathbf{w} \mathbf{w}^{(k), H}\right) \operatorname{vec}\left(\Delta \mathbf{G}\right)+\operatorname{vec}^{H}\left(\Delta \mathbf{G}\right)\left(\mathbf{w}^{*} \mathbf{w}^{(k), T} \otimes \mathbf{v}\right) \Delta \mathbf{h}^{*} \\
&+\Delta \mathbf{h}^{T}\left(\mathbf{w}^{*} \mathbf{w}^{(k), T} \otimes \mathbf{v}^{(k), H}\right) \operatorname{vec}\left(\Delta \mathbf{G}\right)+\operatorname{vec}^{H}\left(\Delta \mathbf{G}\right)\left(\mathbf{w}^{*} \mathbf{w}^{(k), T} \otimes \mathbf{v}^{(k), H}\right) \operatorname{vec}\left(\triangle \mathbf{G}\right) \\
=& {\mathbf{x}}^{H} \mathbf{C} \mathbf{x}+\mathbf{c}_{1}^{H} \mathbf{x}+\mathbf{x}^{H} \mathbf{c}_{2}+c.
\end{aligned}
\end{equation}
\hrulefill
\vspace*{4pt}
\end{figure*}

\begin{appendices}
\section{Proof of Lemma 1}
Let $x$ be a complex scalar variable, we have the first-order Taylor inequality
\begin{equation}
\label{Taylor}
|x|^{2} \geq 2 \operatorname{Re}\left\{x^{*,(k)} x\right\}-x^{*,(k)} x^{(k)}
\end{equation}
for any fixed point $x^{(k)}$. By replacing $x$ and $x^{(k)}$ in \eqref{Taylor} with $\left(\mathbf{h}^{H}+\mathbf{v}^{H} \mathbf{G}\right) \mathbf{w}$ and $\left(\mathbf{h}^{H}+\mathbf{v}^{(k), H} \mathbf{G}\right) \mathbf{w}^{(k)}$, respectively, we have
\begin{align}
\label{Taylor_2}
\notag &\left|\left(\mathbf{h}^{H}+\mathbf{v}^{H} \mathbf{G}\right) \mathbf{w}\right|^{2} \\
&\geq  2 \operatorname{Re}\left\{\left(\mathbf{h}^{H}+\mathbf{v}^{(k),H} \mathbf{G}\right) \mathbf{w}^{(k)} \mathbf{w}^{H}\left(\mathbf{h}+\mathbf{G}^{H} \mathbf{v}\right)\right\} \\
\notag &-\left(\mathbf{h}^{H}+\mathbf{v}^{(k), H} \mathbf{G}\right) \mathbf{w}^{(k)} \mathbf{w}^{(k), H}\left(\mathbf{h}+\mathbf{G}^{H} \mathbf{v}^{(k)}\right) .
\end{align}
The lower bound of \eqref{lemma-1_eq} can be derived from \eqref{Taylor_2} under the channel uncertainty. In particular, we insert $\mathbf{h}=\overline {\mathbf{h}}+\Delta \mathbf{h}$ and $\mathbf{G}=\overline{\mathbf{G}}+\Delta \mathbf{G}$ into the first term on the right hand side of \eqref{Taylor_2}, and then get \eqref{Taylor_3} at the top of the next page. With the similar mathematical transformations, the second terms on the right hand side of \eqref{Taylor_2} under the channel uncertainty can be expressed as
\begin{align}
\notag &\left(\mathbf{h}^{H}+\mathbf{v}^{H} \mathbf{G}\right) \mathbf{w} \mathbf{w}^{(k), H}\left(\mathbf{h}+\mathbf{G}^{H} \mathbf{v}^{(k)}\right) \\
\notag =& \mathbf{x}^{H} \mathbf{C}^{H} \mathbf{x}+\mathbf{c}_{2}^{H} \mathbf{x}+\mathbf{x}^{H} \mathbf{c}_{1}+c^{*} \\
\notag &+\left(\mathbf{h}^{H}+\mathbf{v}^{(k), H} \mathbf{G}\right) \mathbf{w}^{(k)} \mathbf{w}^{(k), H}\left(\mathbf{h}+\mathbf{G}^{H} \mathbf{v}^{(k)}\right) \\
=& \mathbf{x}^{H} \mathbf{Z} \mathbf{x}+\mathbf{z}^{H} \mathbf{x}+\mathbf{x}^{H} \mathbf{z}+z .
\end{align}
\end{appendices}
\\

%======================================================================
%======================================================================
\bibliographystyle{IEEEtran}
\bibliography{IEEEabrv,mybib}

\begin{thebibliography}{10}
\providecommand{\url}[1]{#1}
\csname url@samestyle\endcsname
\providecommand{\newblock}{\relax}
\providecommand{\bibinfo}[2]{#2}
\providecommand{\BIBentrySTDinterwordspacing}{\spaceskip=0pt\relax}
\providecommand{\BIBentryALTinterwordstretchfactor}{4}
\providecommand{\BIBentryALTinterwordspacing}{\spaceskip=\fontdimen2\font plus
\BIBentryALTinterwordstretchfactor\fontdimen3\font minus
  \fontdimen4\font\relax}
\providecommand{\BIBforeignlanguage}[2]{{%
\expandafter\ifx\csname l@#1\endcsname\relax
\typeout{** WARNING: IEEEtran.bst: No hyphenation pattern has been}%
\typeout{** loaded for the language `#1'. Using the pattern for}%
\typeout{** the default language instead.}%
\else
\language=\csname l@#1\endcsname
\fi
#2}}
\providecommand{\BIBdecl}{\relax}
\BIBdecl

\bibitem{UAV_5G}
Y.~Zeng, Q.~Wu, and R.~Zhang, ``Accessing from the sky: A tutorial on {UAV}
  communications for 5{G} and beyond,'' \emph{Proceedings of the IEEE}, vol.
  107, no.~12, pp. 2327--2375, 2019.

\bibitem{UAV_LOS}
D.~Xu, Y.~Sun, D.~W.~K. Ng, and R.~Schober, ``Multiuser {MISO} {UAV}
  communications in uncertain environments with no-fly zones: Robust trajectory
  and resource allocation design,'' \emph{IEEE Trans. Commun.}, vol.~68, no.~5,
  pp. 3153--3172, 2020.

\bibitem{UAV_secure}
B.~Li, Z.~Fei, Y.~Zhang, and M.~Guizani, ``Secure {UAV} communication networks
  over 5{G},'' \emph{IEEE Wirel. Commun.}, vol.~26, no.~5, pp. 114--120, 2019.

\bibitem{UAV_PLS_1}
M.~Cui, G.~Zhang, Q.~Wu, and D.~W.~K. Ng, ``Robust trajectory and transmit
  power design for secure {UAV} communications,'' \emph{IEEE Trans. Veh.
  Technol.}, vol.~67, no.~9, pp. 9042--9046, 2018.

\bibitem{UAV_PLS_2}
H.~Lee, S.~Eom, J.~Park, and I.~Lee, ``{UAV}-aided secure communications with
  cooperative jamming,'' \emph{IEEE Trans. Veh. Technol.}, vol.~67, no.~10, pp.
  9385--9392, 2018.

\bibitem{UAV_PLS_3}
C.~Zhong, J.~Yao, and J.~Xu, ``Secure {UAV} communication with cooperative
  jamming and trajectory control,'' \emph{IEEE Commun. Lett.}, vol.~23, no.~2,
  pp. 286--289, 2019.

\bibitem{UAV_PLS_4}
Y.~Cai, F.~Cui, Q.~Shi, M.~Zhao, and G.~Y. Li, ``Dual-{UAV}-enabled secure
  communications: Joint trajectory design and user scheduling,'' \emph{IEEE J.
  Sel. Areas. Commun.}, vol.~36, no.~9, pp. 1972--1985, 2018.

\bibitem{Deep_UAV_noIRS}
Y.~Zhang, Z.~Mou, F.~Gao, J.~Jiang, R.~Ding, and Z.~Han, ``{UAV}-enabled secure
  communications by multi-agent deep reinforcement learning,'' \emph{IEEE
  Trans. Veh. Technol.}, vol.~69, no.~10, pp. 11\,599--11\,611, 2020.

\bibitem{UAV_PLS_MEC}
Y.~Zhou, C.~Pan, P.~L. Yeoh, K.~Wang, M.~Elkashlan, B.~Vucetic, and Y.~Li,
  ``Secure communications for {UAV}-enabled mobile edge computing systems,''
  \emph{IEEE Trans. Commun.}, vol.~68, no.~1, pp. 376--388, 2020.

\bibitem{UAV_PLS_Vehicle}
T.~Li, J.~Ye, J.~Dai, H.~Lei, W.~Yang, G.~Pan, and Y.~Chen, ``Secure
  {UAV}-to-vehicle communications,'' \emph{IEEE Trans. Commun.}, vol.~69,
  no.~8, pp. 5381--5393, 2021.

\bibitem{UAV_PLS_NOMA}
H.-M. Wang and X.~Zhang, ``{UAV} secure downlink {NOMA} transmissions: A secure
  users oriented perspective,'' \emph{IEEE Trans. Commun.}, vol.~68, no.~9, pp.
  5732--5746, 2020.

\bibitem{UAV_PLS_CR}
Y.~Zhou, F.~Zhou, H.~Zhou, D.~W.~K. Ng, and R.~Q. Hu, ``Robust trajectory and
  transmit power optimization for secure {UAV}-enabled cognitive radio
  networks,'' \emph{IEEE Trans. Commun.}, vol.~68, no.~7, pp. 4022--4034, 2020.

\bibitem{SDOF}
J.~Xie and S.~Ulukus, ``Secure degrees of freedom of one-hop wireless
  networks,'' \emph{IEEE Trans. Inf. Theory}, vol.~60, no.~6, pp. 3359--3378,
  2014.

\bibitem{DEEP_IRS}
Y.~Ge and J.~Fan, ``Beamforming optimization for intelligent reflecting surface
  assisted {MISO}: A deep transfer learning approach,'' \emph{IEEE Trans. Veh.
  Technol.}, vol.~70, no.~4, pp. 3902--3907, 2021.

\bibitem{FullDuplex_IRS}
------, ``Robust secure beamforming for intelligent reflecting surface assisted
  full-duplex {MISO} systems,'' \emph{IEEE Trans. Inf. Forensics Security},
  vol.~17, pp. 253--264, 2022.

\bibitem{IRS_UAV}
X.~Pang, M.~Sheng, N.~Zhao, J.~Tang, D.~Niyato, and K.-K. Wong, ``When {UAV}
  meets {IRS}: Expanding air-ground networks via passive reflection,''
  \emph{IEEE Wirel. Commun.}, vol.~28, no.~5, pp. 164--170, 2021.

\bibitem{UAV_IRS_1}
S.~Fang, G.~Chen, and Y.~Li, ``Joint optimization for secure intelligent
  reflecting surface assisted {UAV} networks,'' \emph{IEEE Wirel. Commun.
  Lett.}, vol.~10, no.~2, pp. 276--280, 2021.

\bibitem{UAV_IRS_2}
L.~Yang, F.~Meng, J.~Zhang, M.~O. Hasna, and M.~D. Renzo, ``On the performance
  of {RIS}-assisted dual-hop {UAV} communication systems,'' \emph{IEEE Trans.
  Veh. Technol.}, vol.~69, no.~9, pp. 10\,385--10\,390, 2020.

\bibitem{IRS_UAV_mmWave}
G.~Sun, X.~Tao, N.~Li, and J.~Xu, ``Intelligent reflecting surface and {UAV}
  assisted secrecy communication in millimeter-wave networks,'' \emph{IEEE
  Trans. Veh. Technol.}, vol.~70, no.~11, pp. 11\,949--11\,961, 2021.

\bibitem{Active_Passive}
Z.~Zhang, L.~Dai, X.~Chen, C.~Liu, and H.~V. Poor, ``Active {RIS} vs. passive
  {RIS}: Which will prevail in 6{G}?'' \emph{arXiv preprint, arXiv:2103.15154},
  2021.

\bibitem{Active_IRS}
R.~Long, Y.-C. Liang, Y.~Pei, and E.~G. Larsson, ``Active reconfigurable
  intelligent surface-aided wireless communications,'' \emph{IEEE Trans. Wirel.
  Commun.}, vol.~20, no.~8, pp. 4962--4975, 2021.

\bibitem{Active_IRS_5}
D.~Xu, X.~Yu, D.~W.~K. Ng, and R.~Schober, ``Resource allocation for active
  {IRS}-assisted multiuser communication systems,'' \emph{arXiv preprint,
  arXiv:2018.13033}, 2021.

\bibitem{Active_IRS_2}
R.~Long, Y.-C. Liang, Y.~Pei, and E.~G. Larsson, ``Active intelligent
  reflecting surface for {SIMO} communications,'' in \emph{Proc. IEEE
  GLOBECOM}, Dec.2020, pp. 1--6.

\bibitem{Active_IRS_3}
C.~You and R.~Zhang, ``Wireless communication aided by intelligent reflecting
  surface: Active or passive?'' \emph{arXiv preprint, arXiv:2106.10963}, 2020.

\bibitem{Active_IRS_4}
L.~Dong, H.-M. Wang, and J.~Bai, ``Active reconfigurable intelligent surface
  aided secure transmission,'' \emph{IEEE Trans. Veh. Technol.}, pp. 1--1,
  2021.

\bibitem{EE_UAV}
H.~Wu, Y.~Wen, J.~Zhang, Z.~Wei, N.~Zhang, and X.~Tao, ``Energy-efficient and
  secure air-to-ground communication with jittering {UAV},'' \emph{IEEE Trans.
  Veh. Technol.}, vol.~69, no.~4, pp. 3954--3967, 2020.

\bibitem{Vib_UAV}
M.~T. Dabiri, M.~Rezaee, V.~Yazdanian, B.~Maham, W.~Saad, and C.~S. Hong,
  ``3{D} channel characterization and performance analysis of {UAV}-assisted
  millimeter wave links,'' \emph{IEEE Trans. Wirel. Commun.}, vol.~20, no.~1,
  pp. 110--125, 2021.

\bibitem{UAV_IRS_3}
S.~Li, B.~Duo, M.~D. Renzo, M.~Tao, and X.~Yuan, ``Robust secure {UAV}
  communications with the aid of reconfigurable intelligent surfaces,''
  \emph{IEEE Trans. Wirel. Commun.}, vol.~20, no.~10, pp. 6402--6417, 2021.

\bibitem{Deep_UAV}
R.~Dong, B.~Wang, and K.~Cao, ``Deep learning driven 3{D} robust beamforming
  for secure communication of {UAV} systems,'' \emph{IEEE Wirel. Commun.
  Lett.}, vol.~10, no.~8, pp. 1643--1647, 2021.

\bibitem{Deep_UAV_2}
X.~Guo, Y.~Chen, and Y.~Wang, ``Learning-based robust and secure transmission
  for reconfigurable intelligent surface aided millimeter wave {UAV}
  communications,'' \emph{IEEE Wirel. Commun. Lett.}, vol.~10, no.~8, pp.
  1795--1799, 2021.

\bibitem{imCSI1}
D.~W.~K. Ng, E.~S. Lo, and R.~Schober, ``Robust beamforming for secure
  communication in systems with wireless information and power transfer,''
  \emph{IEEE Trans. Wirel. Commun.}, vol.~13, no.~8, pp. 4599--4615, 2014.

\bibitem{UAV_IRS_4}
S.~Li, B.~Duo, X.~Yuan, Y.-C. Liang, and M.~Di~Renzo, ``Reconfigurable
  intelligent surface assisted {UAV} communication: Joint trajectory design and
  passive beamforming,'' \emph{IEEE Wirel. Commun. Lett.}, vol.~9, no.~5, pp.
  716--720, 2020.

\bibitem{S_lemma}
E.~Yaz, ``Linear matrix inequalities in system and control theory,''
  \emph{Proceedings of the IEEE}, vol.~86, no.~12, pp. 2473--2474, 1998.

\bibitem{Convex}
Boyd, Vandenberghe, and Faybusovich, ``Convex optimization,'' \emph{IEEE Trans.
  Autom. Control}, vol.~51, no.~11, pp. 1859--1859, 2006.

\bibitem{f_check}
G.~Zhou, C.~Pan, H.~Ren, K.~Wang, M.~D. Renzo, and A.~Nallanathan, ``Robust
  beamforming design for intelligent reflecting surface aided {MISO}
  communication systems,'' \emph{IEEE Wirel. Commun. Lett.}, vol.~9, no.~10,
  pp. 1658--1662, 2020.

\bibitem{Rician}
Q.~Song, F.-C. Zheng, Y.~Zeng, and J.~Zhang, ``Joint beamforming and power
  allocation for {UAV}-enabled full-duplex relay,'' \emph{IEEE Trans. Veh.
  Technol.}, vol.~68, no.~2, pp. 1657--1671, 2019.

\end{thebibliography}

\end{document}